\documentclass[12pt,notitlepage,fleqn]{article}
\usepackage{amssymb}
\usepackage{graphicx}
\usepackage{amsmath}
\usepackage{indentfirst}

\input{tcilatex}

\begin{document}

\begin{center}
\ {\Huge \ Deducing rest masses of quarks with }

{\Huge a three step quantization }

{\LARGE \bigskip }

{\normalsize Jiao Lin Xu}

\ \ \ \ \ \ \ \ \ \ \ \ \ \ \ \ \ \ \ \ \ \ \ \ \ \ \ \ \ \ \ \ \ \ \ \ \ \
\ \ \ \ \ \ \ \ \ \ \ \ \ \ \ \ \ \ \ \ \ \ \ \ \ \ \ \ \ \ \ \ \ \ \ \ \ \
\ \ \ \ \ \ \ \ 

{\small The Center for Simulational Physics, The Department of Physics and
Astronomy}

{\small University of Georgia, Athens, GA 30602, USA}

E- mail: {\small \ jxu@hal.physast.uga.edu}

\bigskip
\end{center}

\textbf{Abstract.}{\small \ Using a three step quantization and
phenomenological formulae, we can deduce the rest masses and intrinsic
quantum numbers (I, S, C, B and Q) of quarks from only one\ unflavored
elementary quark family }$\epsilon ${\small \ with S = C = B = 0 in the
vacuum. Then using sum laws, we can deduce\ the rest masses and intrinsic
quantum numbers of baryons and meson\ from the deduced quarks. The deduced
quantum numbers match experimental results exactly. The deduced rest masses
are consistent with experimental results. This paper predicts some new
quarks \ [d}$_{s}${\small (773), d$_{s}$(1933), }$\text{u}_{c}${\small %
(6073) , }$\text{d}_{b}${\small (9333)], baryons [}$\Lambda _{c} ${\small %
(6699), }$\Lambda _{b}${\small (9959)]\ and mesons [D(6231), B(9502)].\
standard model\ \ }

{\small \ \ \ \ \ \ \ \ \ \ \ \ \ \ \ \ \ \ \ \ \ \ \ \ \ \ \ \ \ \ }

\textbf{1. Introduction}

\ \ \ \ \ \ \ \ \ \ \ \ \ \ \ \ \ \ \ \ \ \ \ \ \ \ \ \ \ \ \ \ 

One hundred years ago, classic physics had already been fully developed.
Most\ physical phenomena could be explained with this physics. Black body
spectrum, however, could not be explained by the physics of that time,
leading Planck to propose a quantization postulate to solve this problem 
\cite{Planck}. The Planck postulate eventually led to quantum mechanics.
Physicists already clearly knew that the black body spectrum was a new
phenomenon outside the applicable area of classic physics. The development
from classic physics to quantum physics depended mainly on new physical
ideas (a quantization postulate) rather than complex mathematics and extra
dimensions of space.

Today we face a similar situation. The standard model \cite{Standard}
\textquotedblleft is in excellent accord with almost all current data.... It
has been enormously successful in predicting a wide range of
phenomena,\textquotedblright\ but\ it cannot deduce the mass spectra of
quarks. So far, no theory has been able to successfully do so. Like black
body spectrum, this mass spectrum may need a new theory outside the standard
model. M. K. Gaillard, P. D. Grannis, and F. J \ Sciulli have already
pointed out \cite{Standard} that the standard model \textquotedblleft is
incomplete... We do not expect the standard model to be valid at arbitrarily
short distances. However, its remarkable success strongly suggest that the
standard model will remain an \ excellent approximation to nature at
distance scales as small as 10$^{-18}$m... high degree of arbitrariness
suggests that a more fundamental theory underlies the standard
model.\textquotedblright\ The mass spectrum of quarks is outside the
applicable area of the standard model. Physicists need to find a new and
more fundamental theory that underlies the standard model. The history of
quantum physics shows that a new physics theory's primary need is new
physical ideas. A three step quantization is the new physical idea. Using
this quantization, we try to deduce the masses of quarks.

Today, physics' foundation is quantum physics (quantum mechanics and quantum
field theory), not classic physics. We work with quantum systems (quarks and
hadrons) that are a level deeper than the system (atoms and molecules) faced
by Planck and Bohr. Therefore, the quantized systems are quantum systems
(not the classic system). If Planck and Bohr got correct quantizations for
atoms and molecules using only one simple quantization, we must use more
steps and more complex quantizations. It is worth emphasizing that deducing
the rest masses and the intrinsic quantum numbers of the quarks using the
three step quantization may be one level deeper than the standard model.
Hopefully, the three step quantization can help physicists discover a more
fundamental theory underlying the standard model \cite{Standard}, just as
the Planck-Bohr quantization did.

\ \ \ \ \ \ \ \ \ \ \ \ \ \ \ \ \ \ \ \ \ \ \ \ \ \ \ \ \ \ \ \ \ \ \ \ 

\textbf{2. The elementary quarks and their free excited quarks}

\ \ \ \ \ \ \ \ \ \ \ \ \ \ \ \ 

\textbf{2.1 The elementary quarks} \textbf{\qquad }

\ \ \ \ \ \ \ \ \ \ \ \ \ \ \ 

We assume that there is only one elementary quark family $\epsilon $ with s
= I = $\frac{1}{2}$ and two isospin states ($\epsilon _{u}$ has I$_{Z}$ = $%
\frac{1}{2}$ and Q = +$\frac{2}{3}$, $\epsilon _{d}$ has I$_{Z}$ = $\frac{-1%
}{2}$ and Q = -$\frac{1}{3}$). For $\epsilon _{u}$ (or $\epsilon _{d}$),\
there are three colored (red, yellow and blue) quarks. Thus, there are six
Fermi elementary quarks in the $\epsilon $ family with S = C = B = 0 in the
vacuum. $\epsilon _{u}$ and $\epsilon _{d}$ have SU(2) symmetry.

As a colored (red, yellow or blue) elementary quark $\epsilon _{u}$ (or $%
\epsilon _{d}$) is excited from the vacuum, its color, electric charge, rest
mass and spin do not change, but it will get energy.\ The free excited
state\ of the elementary quark $\epsilon _{u}$\ is the u-quark with either a
red, yellow or blue color, Q = $\frac{2}{3}$, rest mass m$_{\epsilon
_{u}}^{\ast }$, I = s = $\frac{1}{2}$ and I$_{z}$ = $\frac{1}{2}$ . The free
excited state\ of the elementary quark $\epsilon _{d}$\ is the d-quark with
either a red, yellow or blue color, Q = - $\frac{1}{3},$ rest mass m$%
_{\epsilon _{d}}^{\ast }$, I = s = $\frac{1}{2}$ and I$_{z}$ = -$\frac{1}{2}$%
. Since $\epsilon _{u}$ and $\epsilon _{d}$ have SU(2) symmetry, the u-quark
and the d-quark also have SU(2) symmetry.

\ \ \ \ \ \ \ \ \ \ \ \ \ \ \ \ \ \ \ \ \ 

\textbf{2..2 \ The free motion of excited quark}

\ \ \ \ \ \ \ \ \ \ \ \ \ \ \ \ \ \ \ \ \ \ \ \ \ \ \ \ \ \ \ 

For the excited quark free motion, we generally use the Dirac equation, 
\begin{equation}
\text{i}\hslash \frac{\partial \psi }{\partial \text{t}}\text{=}\frac{%
\hslash c}{i}\text{(}\alpha _{1}\frac{\partial \psi }{\partial x^{1}}\text{+}%
\alpha _{2}\frac{\partial \psi }{\partial x^{2}}\text{+}\alpha _{3}\frac{%
\partial \psi }{\partial x^{3}}\text{) + }\beta m_{\epsilon }^{\ast }\text{C}%
^{2}\psi \text{.}  \label{Dirac}
\end{equation}%
Our purpose, however, is to find rest masses of the excited quarks. The rest
masses are the energy of the excited quark at rest. Corresponding to an
excited quark at rest, the free motion Dirac equation (\ref{Dirac}) reduces 
\cite{Bjorken} to 
\begin{equation}
\text{i}\hslash \frac{\partial \psi }{\partial \text{t}}\text{ = }\beta
m_{\epsilon }^{\ast }\text{c}^{2}\psi .  \label{Rest-Dirac}
\end{equation}%
The equation has four solutions:

\begin{equation*}
\psi ^{1}=e^{-i\frac{\text{mc}^{2}}{\hslash }t}\left[ 
\begin{array}{c}
1 \\ 
0 \\ 
0 \\ 
0%
\end{array}%
\right] ,\psi ^{2}=e^{-i\frac{\text{mc}^{2}}{\hslash }t}\left[ 
\begin{array}{c}
0 \\ 
1 \\ 
0 \\ 
0%
\end{array}%
\right] ,\psi ^{3}=e^{+i\frac{\text{mc}^{2}}{\hslash }t}\left[ 
\begin{array}{c}
0 \\ 
0 \\ 
1 \\ 
0%
\end{array}%
\right] ,\psi ^{1}=e^{+i\frac{\text{mc}^{2}}{\hslash }t}\left[ 
\begin{array}{c}
0 \\ 
0 \\ 
0 \\ 
1%
\end{array}%
\right]
\end{equation*}%
$\psi ^{1}$ and $\psi ^{2}$ correspond to positive energy (quark) and $\psi
^{3}$ and $\psi ^{4}$ correspond to negative energy (antiquark). If we only
consider quark and omit antiquark, we can get the two-component Pauli
equation. Bjorken and Drell wrote \cite{Bjorken}: \textquotedblleft In
particular, we wish to show that they have a sensible nonrelativistic
reduction to the two-component Pauli spin theory.\textquotedblright\ Thus we
can use the low energy limit of the Dirac equation-- Pauli equation to find
the rest masses of quarks.

For single quark free low energy motion, the quark with spin up (s$_{z}$ = $%
\frac{1}{2}$) will have the same energy as the same quark with spin down (s$%
_{z}$ = -$\frac{1}{2}$). Thus, for single free low energy limits, we can
omit the spin of the quark, and the two-component Pauli equation can be
approached by the Schr\"{o}dinger \cite{Schrodinger} equation. The Schr\"{o}%
dinger equation is not to be looked down on. In fact, it is useful in
deducing the rest masses of some baryons \cite{Daliz} and \textquotedblleft
quarkonium\textquotedblright\ mesons \ c$\overline{c}$ and b$\overline{b}$ 
\cite{Martin} and \cite{Kwong}.

When we use the Schr\"{o}dinger equation to approach the Pauli equation, we
cannot forget the static energy of the excited quark. We will deal with this
energy as a constant potential energy (V) at any location. The approximate
Schr\"{o}dinger equation is:

\begin{equation}
\frac{\hslash ^{2}}{\text{2}m_{\epsilon }^{\ast }}\nabla ^{2}\psi \text{ + (}%
\mathbb{E}\text{-V)}\psi \text{ = 0}  \label{Schrodinger}
\end{equation}%
where $m_{\epsilon }^{\ast }$ is the unknown rest mass of the excited quark.
V is the static energy (constant), and it is the minimum excited energy of
an elementary quark from the vacuum. The solution of (\ref{Schrodinger}) is
the eigen wave function and the eigen energy of the free u-quark or the free
d-quark:

\begin{equation}
\begin{tabular}{l}
eigen function $\psi _{\overrightarrow{k}}\text{(}\overrightarrow{\text{r}}%
\text{) }\backsim \text{ exp(i}\overrightarrow{k}\cdot \overrightarrow{r}%
\text{),}$ \\ 
eigen energy $\mathbb{E}\text{ = V +}\frac{\hslash ^{2}}{\text{2}m_{\epsilon
}^{\ast }}\text{[(k}_{1}\text{)}^{2}\text{+(k}_{2}\text{)}^{2}\text{+(k}_{3}%
\text{)}^{2}\text{].}$%
\end{tabular}
\label{Wave+Energy}
\end{equation}%
According to the Quark Model \cite{Quark Model} a proton p = uud and a
neutron n = udd. Omitting electromagnetic mass of quarks, from (\ref%
{Wave+Energy}), at $\overrightarrow{k}$ = 0, we have the rest masses

\begin{eqnarray}
\text{M}_{p} &\text{=}&\text{m}_{u}^{\ast }\text{+m}_{u}^{\ast }\text{+m}%
_{d}^{\ast }\text{ -}\left\vert \text{E}_{bind}\right\vert \approx \text{M}%
_{n}\text{= m}_{u}^{\ast }\text{+m}_{d}^{\ast }\text{+m}_{d}^{\ast }\text{ -}%
\left\vert \text{E}_{bind}\right\vert \text{= 939 Mev }  \label{939} \\
&\rightarrow &\text{m}_{u}^{\ast }\text{ = m}_{d}^{\ast }\text{ = V =\ }%
\frac{1}{3}\text{(939 +}\left\vert \text{E}_{bind}\right\vert \text{) = 313 +%
}\Delta \text{ (Mev)}  \label{313}
\end{eqnarray}%
where E$_{bind}$ is the total binding energy of the three quarks in a
baryon. $\Delta $ represents $\frac{1}{3}\left\vert \text{E}%
_{bind}\right\vert $, and is an unknown large positive constant. Since no
free quark has been found, we assume 
\begin{equation}
\Delta \text{ = }\frac{1}{3}\left\vert \text{E}_{bind}\right\vert \text{ 
\TEXTsymbol{>}\TEXTsymbol{>} M}_{p}\text{.}  \label{Dalta}
\end{equation}%
The free excited u(313+$\Delta $)-quark and d(313+$\Delta $)-quark have
large rest masses (313+$\Delta $) \TEXTsymbol{>}\TEXTsymbol{>} M$_{p}$ = 938
Mev. This is a reason that the Schr\"{o}dinger equation of the low energy
free quark\ is a good approximation of the Dirac equation. The large rest
masses of excited quarks guarantee that the Schr\"{o}dinger equation is a
very good approximation.

Now we deduce the energy bands (quarks) with a three step quantization
method.\ \ 

\ \ \ \ \ \ \ \ \ \ \ \ \ \ \ \ \ \ \ \ \ \ \ \ \ \ \ \ \ \ 

\textbf{3 Deducing energy bands with a three step Quantization}

\ \ \ \ \ \ \ \ \ \ \ \ \ \ \ \ \ \ \ \ \ \ \ \ \ \ \ \ \ \ \ \ 

Today's physics must continue to develop from the standard model into a more
fundamental physics \cite{Standard}. The new theory will deduce the rest
masses and intrinsic quantum numbers (I, S, C, B and Q) of the quarks.
Recall that the development from classic physics to quantum physics began
with the Planck-Bohr quantization. We try to use a three step quantization
method to start the long time procedure.

\ \ \ \ \ \ \ \ \ \ \ \ \ \ \ \ \ \ \ \ \ \ \ \ \ \ \ \ \ \ \ \ \ \ \ \ \ \
\ \ \ \ \ \ \ \ \ \ 

\textbf{3.1 Recall Planck and Bohr's works}

\ \ \ \ \ \ \ \ \ \ \ \ \ \ \ \ \ \ \ 

Planck's \cite{Planck} energy quantization postulate states that
\textquotedblleft any physical entity whose single `coordinate'\ execute
simple harmonic oscillations (i.e., is a sinusoidal function of time) can
possess only total energy $\varepsilon $ which satisfy the relation $%
\varepsilon =nh\nu $, n = 0, 1, 2, 3, .... where $\nu $ is the frequency of
the oscillation and h is a universal constant.\textquotedblright\ Planck
selects reasonable energy from a continuous energy spectrum. \ \ \ \ \ \ \ \
\ \ \ \ 

Bohr's \cite{Bohr} orbit quantization tells us that \textquotedblleft an
electron in an atom moves in a circular orbit about the nucleus...obeying
the laws of classical mechanics. But instead of the infinity of orbits which
would be possible in classical mechanics, it is only possible for an
electron to move in an orbit for which its orbital angular momentum L is an
integral multiple of Planck's constant h, divided by 2$\pi $%
.\textquotedblright\ Using the quantization condition (L = $\frac{nh}{2\pi }$%
, \ n = 1, 2, 3, ... ), Bohr selects reasonable orbits from infinite orbits.

Drawing from these great physicists' works, we find that the most important
law is to use quantized conditions and symmetries (as circular orbit) to
select reasonable energy levels from a continuous energy spectrum. \ \ 

\ \ \ \ \ \ \ \ \ \ 

\textbf{3.2 Three step Quantization}

\ \ \ \ \ \ \ \ \ \ \ \ \ \ \ \ \ \ \ 

In order to get the short-lived quarks, we quantize the free motion (quantum
plane wave function of the Schr\"{o}dinger equation) of an excited quark (%
\ref{Wave+Energy}) to select energy bands from the continuous energy.\ The
energy bands will correspond to short-lived quarks.

\ \ \ \ \ \ \ \ \ \ \ \ \ \ \ \ \ \ \ \ \ \ \ \ \ \ 

3.2.1 The first step quantizing condition

\ \ \ \ \ \ \ \ \ \ \ \ 

For free motion of an excited quark with continuous energy (\ref{Wave+Energy}%
),\ we assume the wave vector $\overrightarrow{k}$ has the symmetries of the
regular rhombic dodecahedron in $\overrightarrow{k}$-space (see Fig. 1).

\ \ \ \ \ \ \ \ \ \ \ \ \ \ \ \ \ \ \ \ \ \ \ \ \ \ \ \ \ \ \ \ \ \ \ \ \ \
\ \ \ 

\ \ \ \ \ \ \ \ \ \ \ \ \ \ \ \ \ \ \ \ \ \ \ \ \ \ \ \ \ \ \ \ \ \ \ \ \ 

\ \ \ \ \ \ \ \ \ \ \ \ \ \ \ 

\begin{figure}[h]
\vspace{5.8in} \includegraphics{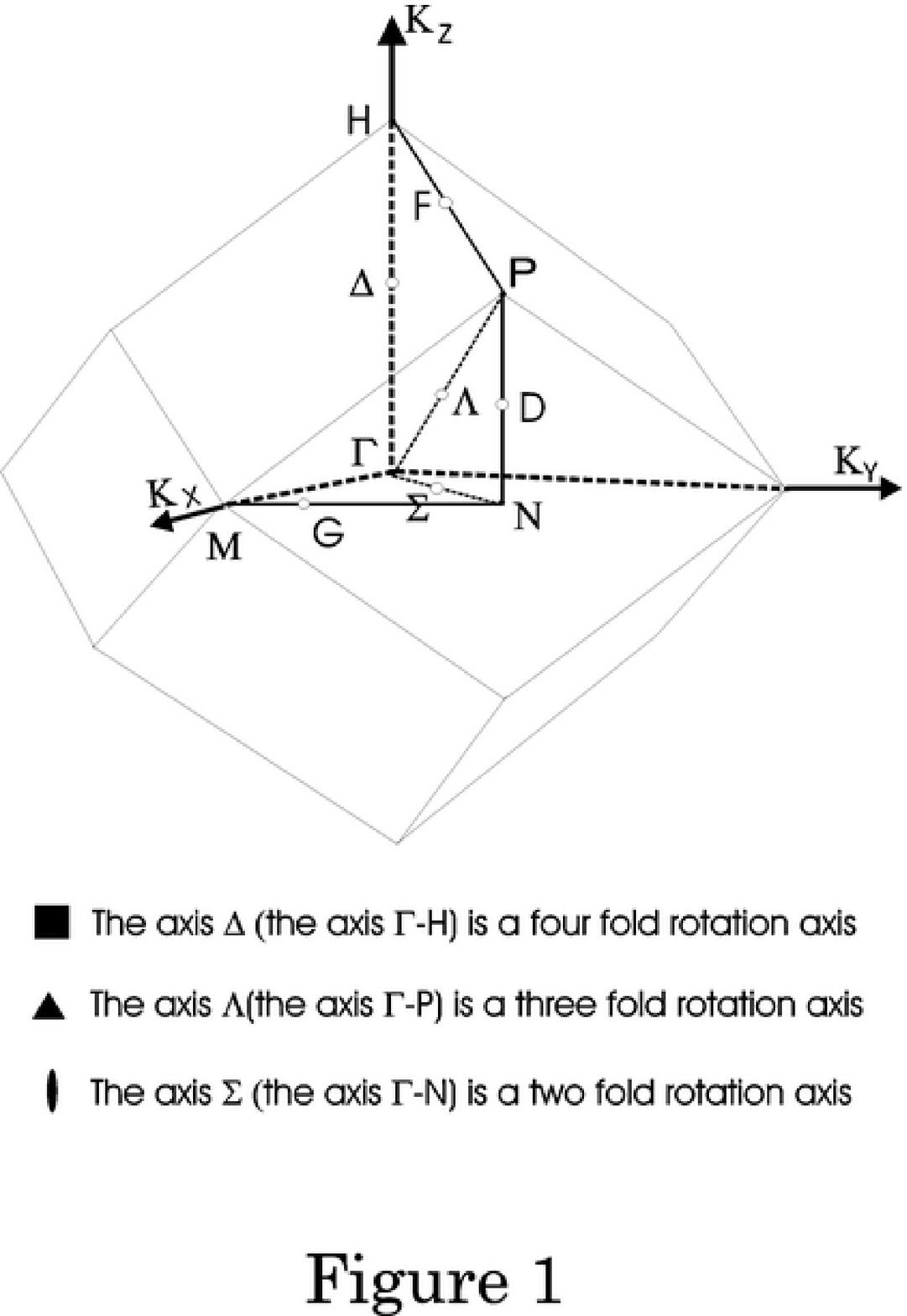} \label{Fig1}
\caption{{\protect\small The regular rhombic dodecahedron. The symmetry
points and axes are indicated.}}
\end{figure}

\bigskip\ \ \ \ \ \ \ \ \ \ \ \ \ \ \ \ \ \ \ \ \ \ \ \ \ \ \ \ \ \ \ \ \ \
\ \ \ \ \ \ \ \ \ \ \ \ \ \ \ \ \ \ \ \ \ \ \ \ \ \ 

We assume that the axis $\Gamma $-H in Fig.1 has length $\frac{2\pi }{a}$
with an unknown constant a. The first step quantizing conditions are: 
\begin{equation}
\begin{tabular}{l}
k$_{1}$ = $\frac{2\pi }{a}$(n$_{1}$-$\xi $), \\ 
k$_{2}$ = $\frac{2\pi }{a}$(n$_{2}$-$\eta $), \\ 
k$_{3}$ = $\frac{2\pi }{a}$(n$_{3}$-$\zeta $).%
\end{tabular}
\label{K(1,2,3)}
\end{equation}%
Putting (\ref{K(1,2,3)}) into (\ref{Wave+Energy}), we get (\ref{BandW})$\ $%
and (\ref{E(nk)}):

\begin{eqnarray}
\psi _{\overrightarrow{k}}\text{(}\overrightarrow{\text{r}}\text{) }
&\thickapprox &\text{ exp}\frac{2\pi i}{a}\text{[(n}_{1}\text{-}\xi \text{)x
+ (n}_{2}\text{ - }\eta \text{)y+(n}_{3}\text{ - }\zeta \text{)z],}
\label{BandW} \\
\mathbb{E}\text{(}\vec{k}\text{,}\vec{n}\text{) } &\text{=}&\text{313 + }%
\Delta \text{ + }\alpha \text{[(n}_{1}\text{-}\xi \text{)}^{2}\text{+(n}_{2}%
\text{-}\eta \text{)}^{2}\text{+(n}_{3}\text{-}\zeta \text{)}^{2}\text{]} 
\notag \\
&\text{=}&\text{313 + }\Delta \text{ + 360 E(}\overrightarrow{\kappa }\text{,%
}\overrightarrow{n}\text{) (Mev).}  \label{E(nk)}
\end{eqnarray}%
where $\alpha $ =$\frac{\text{h}^{2}}{\text{2m}_{\epsilon }^{\ast }\text{a}%
^{2}}$ = 360 Mev (\ref{360}). E($\overrightarrow{\kappa }$,$\overrightarrow{n%
}$) = [(n$_{1}$-$\xi $)$^{2}$+(n$_{2}$-$\eta $)$^{2}$+(n$_{3}$-$\zeta $)$%
^{2} $]. For $\overrightarrow{n}$ = (n$_{1}$, n$_{2}$, n$_{3}$), n$_{1}$, n$%
_{2}$ and n$_{3}$ are $\pm \func{integer}$s and zero. The $\overrightarrow{%
\kappa } $ = ($\xi $, $\eta $, $\zeta $) has the symmetries of a regular
rhombic dodecahedron. In order to deduce the short-lived quarks, we must
further quantize the $\overrightarrow{n}$ and $\overrightarrow{\kappa }$ as
follows.

\ \ \ \ \ \ \ \ \ \ \ \ \ \ \ \ \ \ \ \ \ \ \ \ \ \ \ \ \ \ \ 

3.2.2 The second step quantization

\ \ \ \ \ \ \ \ \ \ \ \ \ \ \ \ \ \ \ \ \ \ \ \ \ \ \ \ \ \ \ \ \ \ \ \ \ \
\ \ \ \ \ \ \ \ \ \ \ \ 

We will quantize the $\overrightarrow{n}$ = (n$_{1}$, n$_{2}$, n$_{3}$)
values further. If we assume n$_{1}$ = \textit{l}$_{2}$ \textit{+ l}$_{3}$, n%
$_{2}$ =\textit{\ l}$_{3}$ \textit{+ l}$_{1}$ and n$_{3}$ =\textit{\ l}$_{1}$
\textit{+ l}$_{2},$ so that 
\begin{equation}
\begin{tabular}{l}
\textit{l}$_{1}$ = $\frac{1}{2}$(-n$_{1}$ + n$_{2}$ + n$_{3}$) \\ 
\textit{l}$_{2}$ = $\frac{1}{2}$(+n$_{1}$ - n$_{2}$ + n$_{3}$) \\ 
\textit{l}$_{3}$ = $\frac{1}{2}$(+n$_{1}$ + n$_{2}$ - n$_{3}$).%
\end{tabular}
\label{l-n}
\end{equation}%
The second step quantizing condition is that only those values of $%
\overrightarrow{n}$ = (n$_{1}$, n$_{2}$, n$_{3}$) are allowed that make $%
\overrightarrow{l}$ = \textit{(l}$_{1}$\textit{, l}$_{2}$\textit{, l}$_{3}$\
) an integer vector. This is a second step quantization for $\vec{n}$
values. For example, $\vec{n}$ \ cannot take the values (1, 0, 0) or (1, 1,
-1), but can take (0, 0, 2) and (1, -1, 2). From E($\overrightarrow{\kappa }$%
,$\overrightarrow{n}$) = [(n$_{1}$-$\xi $)$^{2}$ +(n$_{2}$-$\eta $)$^{2}$ +(n%
$_{3}$-$\zeta $)$^{2}$], we can give a definition of the equivalent $%
\overrightarrow{n}$: for $\overrightarrow{\kappa }$ = ($\xi $, $\eta ,$ $%
\varsigma )$ = 0, all $\overrightarrow{n}$ values that give the same E($%
\overrightarrow{\kappa }$,$\overrightarrow{n}$) value are equivalent
n-values. We show the low level equivalent $\overrightarrow{n}$-values that
satisfy condition (\ref{l-n}) in the following list (\ref{nnn}) (note $%
\overline{\text{n}_{i}}$ = - n$_{i}$):\ \ 

\begin{equation}
\begin{tabular}{|l|}
\hline
{\small E(}$\overrightarrow{n}${\small ,0) = 0\ : (0, 0, 0) \ \ \ \ \ \ \ \
\ \ Notes: }[$\overline{\text{{\small 1}}}${\small 12 }$\equiv $ (-1,1,2)
and $\overline{\text{{\small 1}}}${\small 1}$\overline{\text{{\small 2}}}%
\equiv $ (-1,1,-2)] \\ \hline
{\small E(}$\overrightarrow{n}${\small ,0) = 2\ :}$\ ${\small (101, }$%
\overline{\text{{\small 1}}}${\small 01, 011, 0}$\overline{\text{{\small 1}}}
${\small 1, 110, 1}$\overline{\text{{\small 1}}}${\small 0, }$\overline{%
\text{{\small 1}}}${\small 10, }$\overline{\text{{\small 1}}}\overline{\text{%
{\small 1}}}${\small 0, 10}$\overline{\text{{\small 1}}}${\small , }$%
\overline{\text{{\small 1}}}${\small 0}$\overline{\text{{\small 1}}}${\small %
, 01}$\overline{\text{{\small 1}}}${\small , 0}$\overline{\text{{\small 1}}}%
\overline{\text{{\small 1}}}${\small )} \\ \hline
{\small E(}$\overrightarrow{n}${\small ,0) = 4\ :\ \ (002, 200}, {\small 200}%
$\text{, }\overline{\text{{\small 2}}}\text{{\small 00}, }${\small 0}$%
\overline{\text{{\small 2}}}${\small 0, 00}$\overline{\text{{\small 2}}}$%
{\small )} \\ \hline
{\small E(}$\overrightarrow{n}${\small ,0) = 6:} 
\begin{tabular}{l}
{\small 112, 211, 121, }$\overline{\text{{\small 1}}}${\small 21,}$\overline{%
\text{{\small 1}}}${\small 12, 2}$\overline{\text{{\small 1}}}${\small 1}, 
{\small 1}$\overline{\text{{\small 1}}}${\small 2, 21}$\overline{\text{%
{\small 1}}}${\small ,12}$\overline{\text{{\small 1}}}${\small ,}$\overline{%
\text{{\small 2}}}${\small 11, 1}$\overline{\text{{\small 2}}}${\small 1, 11}%
$\overline{\text{{\small 2}}}$,{\small \ } \\ 
$\overline{\text{{\small 11}}}${\small 2, }$\overline{\text{{\small 1}}}$%
{\small 2}$\overline{\text{{\small 1}}}$,{\small \ 2}$\overline{\text{%
{\small 11}}}${\small , }$\overline{\text{{\small 21}}}${\small 1}, $%
\overline{\text{{\small 12}}}${\small 1, 1}$\overline{\text{{\small 12}}}$%
{\small , 1}$\overline{\text{{\small 21}}}${\small ,}$\overline{\text{%
{\small 1}}}${\small 1}$\overline{\text{{\small 2}}}${\small ,}$\overline{%
\text{{\small 2}}}$1$\overline{\text{{\small 1}}}$,{\small \ }$\overline{%
\text{{\small 211}}}${\small , }$\overline{\text{{\small 121}}},${\small \ }$%
\overline{\text{{\small 112}}},$%
\end{tabular}
\\ \hline
\end{tabular}
\label{nnn}
\end{equation}

$\ \ \ \ \ \ \ \ \ \ \ \ \ \ \ \ \ \ \ \ \ \ \ \ \ \ $\ \ \ \ \ \ \ \ \ \ \
\ \ 

3.2.3 The third step Quantization

\ \ \ \ \ \ \ \ \ \ \ \ \ \ \ \ \ \ \ \ \ \ \ 

The\ vector $\overrightarrow{\kappa }$ = ($\xi $, $\eta $, $\zeta $) in (\ref%
{BandW}) and\ (\ref{E(nk)}) has the symmetries of the regular rhombic
dodecahedron in k-space (see Fig. 1). From Fig. 1, we can see that there are
four kinds of symmetry points ($\Gamma $, H, P and N) and six kinds of
symmetry axes ($\Delta $, $\Lambda $, $\Sigma $, D, F and G) in the regular
rhombic dodecahedron. The coordinates ($\xi $, $\eta $, $\varsigma $) of the
symmetry points and axes are:

\begin{equation}
\overrightarrow{\kappa }_{\Gamma }=(0,0,0),\overrightarrow{\kappa }_{\text{H}%
}=(0,0,1),\overrightarrow{\kappa }_{\text{p}}\text{ }\text{= (}\frac{\text{1}%
}{\text{2}}\text{, }\frac{\text{1}}{\text{2}}\text{, }\frac{\text{1}}{\text{2%
}}\text{), }\overrightarrow{\kappa }_{\text{N}}\text{ }\text{= (}\frac{\text{%
1}}{\text{2}}\text{, }\frac{\text{1}}{\text{2}}\text{, }0\text{).}
\label{S-Point}
\end{equation}%
\begin{equation}
\begin{tabular}{ll}
$\overrightarrow{\kappa }_{\Delta }\text{ }\text{= (0, 0, }\zeta \text{),\ \
0}\leq \zeta \text{ }\leq \text{1; }$ & $\overrightarrow{\kappa }_{\Lambda }%
\text{ = (}\xi \text{, }\xi \text{, }\xi \text{), \ 0}\leq \xi \leq \frac{%
\text{1}}{\text{2}}\text{;}$ \\ 
$\overrightarrow{\kappa }_{\Sigma }\text{{} }\text{= (}\xi \text{, }\xi 
\text{, 0), \ 0}\leq \xi \text{ }\leq \frac{\text{1}}{\text{2}}\text{;}$ & $%
\overrightarrow{\kappa }_{\text{D}}\text{ }\text{= (}\frac{\text{1}}{\text{2}%
}\text{, }\frac{\text{1}}{\text{2}}\text{, }\xi \text{), \ 0}\leq \xi \leq 
\frac{\text{1}}{\text{2}}\text{;}$ \\ 
$\overrightarrow{\kappa }_{\text{G}}\text{ }\text{= (}\xi \text{, 1-}\xi 
\text{, 0), \ }\frac{\text{1}}{\text{2}}\leq \xi \text{ }\leq \text{1;}$ & $%
\overrightarrow{\kappa }_{\text{F}}\text{ = (}\xi \text{, }\xi \text{, 1-}%
\xi \text{), \ 0}\leq \xi \leq \frac{\text{1}}{\text{2}}\text{.}$%
\end{tabular}
\label{Sym-Axes}
\end{equation}%
The third step quantizing condition is that the coordinates ($\xi $, $\eta $%
, $\zeta $) of $\ \overrightarrow{\kappa }$ in (\ref{E(nk)}) only take the
coordinate values of the six symmetry axes (\ref{Sym-Axes}).

\ \ \ \ \ \ \ \ \ \ \ \ \ \ \ \ 

\textbf{3.3 Energy bands}

\ \ \ \ \ \ \ \ \ \ \ \ \ \ \ \ \ \ \ \ \ \ \ \ \ \ \ \ 

From the low energy free wave motion of a excited elementary quark $\epsilon 
$ with a continuous energy spectrum \{$\mathbb{E}$ = V +$\frac{\hslash ^{2}}{%
2m}$[(k$_{1}$)$^{2}$+(k$_{2}$)$^{2}$+(k$_{3}$)$^{2}$] (\ref{Wave+Energy})\},
using the three step quantization, we obtain a new energy formula \{$\mathbb{%
E}$($\vec{k}$,$\vec{n}$) =313 + $\Delta $ + $\alpha $[(n$_{1}$-$\xi $)$^{2}$%
+(n$_{2}$-$\eta $)$^{2}$+(n$_{3}$-$\zeta $)$^{2}$] (\ref{E(nk)})\} with
quantized $\vec{n}$ values of (\ref{nnn}) and $\vec{k}$ values of (\ref%
{Sym-Axes}). The energy (\ref{E(nk)}) with a $\overrightarrow{n}$\ = (n$_{1}$%
, n$_{2}$, n$_{3}$) of (\ref{nnn}) and a $\vec{k}$ = ($\xi $, $\eta $, $%
\varsigma $) of (\ref{Sym-Axes}) forms an energy band. The formula (\ref%
{E(nk)}) with quantized $\vec{n}$ of (\ref{nnn}) and $\vec{k}$ of (\ref%
{Sym-Axes}) is the formula that can deduced all energy bands.

\ \ \ \ \ \ \ \ \ \ \ \ \ \ \ \ \ \ \ \ 

\textbf{3.4 Deducing energy bands}

\ \ \ \ \ \ \ \ \ \ \ \ \ \ \ \ \ \ \ \ \ \ \ 

After getting (\ref{E(nk)}), (\ref{nnn}) and (\ref{Sym-Axes}), we can deduce
low energy bands of the six symmetry axes. As an example, we will deduce the
single energy bands of the $\Delta $-axis.\ For the $\Delta $-axis, $%
\overrightarrow{\kappa }_{\Delta }$\ = (0, 0, $\zeta $) from (\ref{Sym-Axes}%
). Putting $\overrightarrow{\kappa }_{\Delta }$\ = (0, 0, $\zeta $) into (%
\ref{E(nk)}), we get $\mathbb{E}_{\Delta }${\small (}$\vec{k}${\small ,}$%
\vec{n}${\small ) }= 313+$\Delta $+360[(n$_{1}$)$^{2}$+(n$_{2}$)$^{2}$+(n$%
_{3}$-$\zeta $)$^{2}$]. For point$\Gamma ,\overrightarrow{\kappa }_{\Gamma }$%
\ = (0, 0, $0$) from (\ref{S-Point}), E$_{\Gamma }$($\overrightarrow{\kappa }
$,$\overrightarrow{n}$) = (n$_{1}$)$^{2}$+(n$_{2}$)$^{2}$+(n$_{3}$)$^{2}$.
For point-H$,\overrightarrow{\kappa }_{\text{H}}$\ = (0, 0, 1) from (\ref%
{S-Point}), E$_{\text{H}}$($\overrightarrow{\kappa }$,$\overrightarrow{n}$)
= (n$_{1}$)$^{2}$+(n$_{2}$)$^{2}$+(n$_{3}$-1)$^{2}$. 
\begin{equation}
\begin{tabular}{l}
T$\text{he }\Delta \text{-axis, }\mathbb{E}_{\Delta }\text{{\small (}}\vec{k}%
\text{{\small ,}}\vec{n}\text{{\small )}=313+}\Delta \text{+360[(n}_{1}\text{%
)}^{2}\text{+(n}_{2}\text{)}^{2}\text{+(n}_{3}\text{-}\zeta \text{)}^{2}%
\text{]}$ \\ 
the $\Gamma $-$\text{point, }\overrightarrow{\kappa }_{\Gamma }\text{\
=(0,0,0), E}_{\Gamma }\text{(}\overrightarrow{\kappa }\text{,}%
\overrightarrow{n}\text{)=(n}_{1}\text{)}^{2}\text{+(n}_{2}\text{)}^{2}\text{%
+(n}_{3}\text{)}^{2}$ \\ 
the H-$\text{point, }\overrightarrow{\kappa }_{\text{H}}\text{\ =(0,0,1), E}%
_{\text{H}}\text{(}\overrightarrow{\kappa }\text{,}\overrightarrow{n}\text{%
)=(n}_{1}\text{)}^{2}\text{+(n}_{2}\text{)}^{2}\text{+(n}_{3}\text{-1)}^{2}%
\text{.}$%
\end{tabular}
\label{D-1}
\end{equation}%
Putting $\text{(n}_{1}\text{, n}_{2}\text{, n}_{3}${\small ) }values of the
single bands of the $\Delta $-axis (Table B1 of \cite{0502091}) into $%
\mathbb{E}_{\Delta }$($\vec{k}$,$\vec{n}$), E$_{\Gamma }$($\overrightarrow{%
\kappa }$,$\overrightarrow{n}$) and E$_{\text{H}}$($\overrightarrow{\kappa }$%
,$\overrightarrow{n}$) (\ref{D-1}), we can find energy bands as shown in
Table 1:

\ \ \ \ \ \ \ 

\begin{tabular}{|l|}
\hline
\ \ \ \ \ Table 1\ The Single Energy Bands of the $\Delta $-Axis (the $%
\Gamma $-H axis)\  \\ \hline
$%
\begin{tabular}{|l|l|l|l|l|}
\hline
$\text{(n}_{1}\text{,n}_{2}\text{,n}_{3}${\small )} & $\text{E(}%
\overrightarrow{\kappa }\text{,}\overrightarrow{n}\text{)}_{Start}$ & minim.
E & $\mathbb{E}${\small (}$\vec{k}${\small ,}$\vec{n}${\small ) (Band)} & $%
\text{E(}\overrightarrow{\kappa }\text{,}\overrightarrow{n}\text{)}_{end}$
\\ \hline
{\small (0, 0, 0)} & $\text{E}_{\Gamma }\text{=0}$ & 313+$\Delta $ & 313+$%
\Delta $+$\zeta ^{2}$ & $\text{E}_{H}\text{=1}$ \\ \hline
{\small (0, 0, 2) } & $\text{E}_{H}\text{=1}$ & {\small 673}+$\Delta $ & 313+%
$\Delta $+(2-$\zeta $)$^{2}$ & $\text{E}_{\Gamma }\text{=4}$ \\ \hline
{\small (0, 0, }$\overline{2}${\small ) } & $\text{E}_{\Gamma }\text{=4}$ & 
{\small 1753}+$\Delta $ & 313+$\Delta $+(2+$\zeta $)$^{2}$ & $\text{E}_{H}%
\text{=9}$ \\ \hline
{\small (0, 0, 4)}$\text{ }$ & $\text{E}_{H}\text{=9}$ & {\small 3553}+$%
\Delta $ & 313+$\Delta $+(4-$\zeta $)$^{2}$ & $\text{E}_{\Gamma }\text{=16}$
\\ \hline
{\small (0, 0, }$\overline{\text{{\small 4}}}${\small )} & $\text{E}_{\Gamma
}\text{=16}$ & {\small 6073}+$\Delta $ & 313+$\Delta $+(4+$\zeta $)$^{2}$ & $%
\text{E}_{H}\text{=25}$ \\ \hline
(0, 0, 6) & $\text{E}_{H}\text{=25}$ & {\small 9313}+$\Delta $ & 313+$\Delta 
$+(6-$\zeta $)$^{2}$ & $\text{E}_{\Gamma }\text{=36}$ \\ \hline
& ... & ... & ... & ... \\ \hline
\end{tabular}%
\ \ $ \\ \hline
$\text{E(}\overrightarrow{\kappa }\text{,}\overrightarrow{n}\text{)}_{Start}$%
{\small is the value of }$\text{E(}\overrightarrow{\kappa }\text{,}%
\overrightarrow{n}\text{) at the start point of the energy band}$ \\ \hline
$\text{E(}\overrightarrow{\kappa }\text{,}\overrightarrow{n}\text{)}_{end}$%
{\small is the value of }$\text{E(}\overrightarrow{\kappa }\text{,}%
\overrightarrow{n}\text{) at the end point of the energy band}$ \\ \hline
\end{tabular}

\ \ \ \ \ \ \ \ \ \ \ \ \ \ \ \ \ \ \ \ \ \ \ \ \ \ \ \ \ \ \ \ \ \ \ \ \ \
\ \ \ \ \ 

Similarly, we can deduce the single energy bands of the $\Sigma $-axis. For
the $\Sigma $-axis, $\overrightarrow{\kappa }_{\Sigma }$= ($\xi $, $\xi $, $%
0 $). Putting the $\overrightarrow{\kappa }$ into (\ref{E(nk)}), we have 
{\small \ }$\mathbb{E}_{\Sigma }${\small (}$\vec{k}${\small ,}$\vec{n}$%
{\small )} =313+ $\Delta $ + $360$[(n$_{1}$-$\xi $)$^{2}$+(n$_{2}$-$\xi $)$%
^{2}$+(n$_{3}$)$^{2}$] $\text{\ 0}\leq \zeta \leq \frac{\text{1}}{2}$. For
point N$,$ $\overrightarrow{\kappa }_{\text{N}}$\ = ($\frac{1}{2}$, $\frac{1%
}{2}$, 0) from (\ref{S-Point}), E$_{\text{N}}$($\overrightarrow{\kappa }$,$%
\overrightarrow{n}$) = (n$_{1}$-$\frac{1}{2}$)$^{2}$ +(n$_{2}$-$\frac{1}{2}$)%
$^{2}$ +(n$_{3}$)$^{2}$.%
\begin{equation}
\begin{tabular}{l}
$\mathbb{E}_{\Sigma }${\small (}$\vec{k}${\small ,}$\vec{n}${\small )} =313+ 
$\Delta $ + $360$[(n$_{1}$-$\xi $)$^{2}$+(n$_{2}$-$\xi $)$^{2}$+(n$_{3}$)$%
^{2}$] \\ 
$\overrightarrow{\kappa }_{\text{N}}$\ = ($\frac{1}{2}$, $\frac{1}{2}$, 0), E%
$_{\text{N}}$($\overrightarrow{\kappa }$,$\overrightarrow{n}$) = (n$_{1}$-$%
\frac{1}{2}$)$^{2}$+(n$_{2}$-$\frac{1}{2}$)$^{2}$+(n$_{3}$)$^{2}$ \\ 
$\overrightarrow{\kappa }_{\Gamma }$\ = (0, 0, 0), E$_{\text{N}}$($%
\overrightarrow{\kappa }$,$\overrightarrow{n}$) = (n$_{1}$)$^{2}$+(n$_{2}$)$%
^{2}$+(n$_{3}$)$^{2}$ {\small (\ref{D-1})}%
\end{tabular}
\label{S-1}
\end{equation}%
Putting $\text{(n}_{1}\text{,n}_{2}\text{,n}_{3}${\small ) }values of the
single bands of the $\Sigma $-axis (Table B2 of \cite{0502091}) into $%
\mathbb{E}$($\vec{k}$,$\vec{n}$), {\small E}$_{\text{N}}${\small (}$%
\overrightarrow{\kappa }${\small ,}$\overrightarrow{n}${\small ) and E}$%
_{\Gamma }${\small (}$\overrightarrow{\kappa }${\small ,}$\overrightarrow{n}$%
{\small ) (\ref{S-1}), }we can deduce energy bands as shown in Table 2

\ \ \ \ \ \ \ \ \ \ \ \ \ \ \ \ \ \ \ \ \ \ \ \ \ \ \ \ \ \ \ 

\begin{tabular}{|l|}
\hline
\ \ \ \ Table 2\ The Single Energy Bands of the $\Sigma $-Axis (the $\Gamma $%
-N axis)\ $\text{ }$ \\ \hline
$%
\begin{tabular}{|l|l|l|l|l|}
\hline
$\text{\ n}_{1}\text{n}_{2}\text{n}_{3}$ & $\text{E(}\overrightarrow{\kappa }%
\text{,}\overrightarrow{n}\text{)}_{Start}$ & minim.E & $\mathbb{E}_{\Sigma
} ${\small =313+}+$\Delta $+360$\text{E(}\overrightarrow{\kappa }\text{,}%
\overrightarrow{n}\text{)}$ & $\text{E(}\overrightarrow{\kappa }\text{,}%
\overrightarrow{n}\text{)}_{end}$ \\ \hline
{\small (0, 0, 0)} & $\text{E}_{\Gamma }\text{=0}$ & {\small 313}+ $\Delta $
& 313+$\Delta $+720$\xi ^{2}$ & $\text{E}_{N}\text{=}\frac{\text{1}}{2}$ \\ 
\hline
$(\text{1,1,0})$ & $\text{E}_{N}\text{=}\frac{\text{1}}{2}$ & $\text{493}$+ $%
\Delta $ & 313+$\Delta $+720(1-$\xi $)$^{2}$ & $\text{E}_{\Gamma }\text{=2}$
\\ \hline
$(\text{-1,-1,0})$ & $\text{E}_{\Gamma }\text{=2}$ & $\text{1033}$+$\Delta $
& 313+$\Delta $+720(1+$\xi $)$^{2}$ & $\text{E}_{N}\text{=}\frac{\text{9}}{2}
$ \\ \hline
$(\text{2,2,0})$ & $\text{E}_{N}\text{=}\frac{\text{9}}{2}$ & $\text{1933}$+$%
\Delta $ & 313+$\Delta $+720(2-$\xi $)$^{2}$ & $\text{E}_{\Gamma }\text{=8}$
\\ \hline
$(\text{-2,-2,0})$ & $\text{E}_{\Gamma }\text{=8}$ & $\text{3193}$+$\Delta $
& 313+$\Delta $+720(2+$\xi $)$^{2}$ & $\text{E}_{N}\text{=}\frac{\text{25}}{2%
}$ \\ \hline
$(\text{3,3,0})$ & $\text{E}_{N}\text{=}\frac{\text{25}}{2}$ & $\text{4813}$+%
$\Delta $ & 313+$\Delta $+720(3-$\xi $)$^{2}$ & $\text{E}_{\Gamma }\text{=18}
$ \\ \hline
{\small (-}3,-3,0{\small )} & $\text{E}_{\Gamma }\text{=18}$ & $\text{6793}$+%
$\Delta $ & 313+$\Delta $+720(3+$\xi $)$^{2}$ & $\text{E}_{N}\text{=}\frac{%
\text{49}}{2}$ \\ \hline
(4, 4, 0) & $\text{E}_{N}\text{=}\frac{\text{49}}{2}$ & $\text{9133}$+$%
\Delta $ & 313+$\Delta $+720(4-$\xi $)$^{2}$ & $\text{E}_{\Gamma }\text{=32}$
\\ \hline
\end{tabular}%
\ $ \\ \hline
$\text{E(}\overrightarrow{\kappa }\text{,}\overrightarrow{n}\text{)}_{Start}$%
{\small is the value of }$\text{E(}\overrightarrow{\kappa }\text{,}%
\overrightarrow{n}\text{) at the start point of the energy band}$ \\ \hline
$\text{E(}\overrightarrow{\kappa }\text{,}\overrightarrow{n}\text{)}_{end}$%
{\small is the value of }$\text{E(}\overrightarrow{\kappa }\text{,}%
\overrightarrow{n}\text{) at the end point of the energy band}$ \\ \hline
\end{tabular}

\ \ \ \ \ \ \ \ \ \ \ \ \ \ \ \ \ \ \ \ \ \ \ \ \ \ \ \ \ \ \ \ \ \ \ \ \ \
\ \ \ \ 

Following the two examples in Tables 1 and 2, we can deduce all low energy
bands of the six axes using the formulae (\ref{E(nk)}), (\ref{nnn}) and (\ref%
{Sym-Axes}) (see Appendix B of \cite{0502091})

From these energy bands, we can deduce quarks using phenomenological
formulae.

\ \ \ \ \ \ \ \ \ \ \ \ \ \ \ \ \ \ \ \ \ \ \ \ \ \ \ \ \ \ \ \ \ 

\textbf{4 The phenomenological formulae}{\LARGE \ }

\ \ \ \ \ \ \ \ \ \ \ \ \ \ \ \ \ \ \ \ \ \ \ \ \ \ \ \ \ \ \ \ \ \ \ \ \ \
\ \ \ \ \ \ \ \ \ \ \ \ \ \ \ \ \ \ \ \ \ \ \ \ \ \ \ \ \ \ \ \ \ \ \ \ \ \
\ \qquad \qquad \qquad \qquad\ \ \ \ \ \ \ \ \ \ \ \ \ \ \ \ \ \ \ \ \ \ \ \
\ \ \ \ \ \ \ \ \ \ \ \ \ \ \ \ \ \ \ \ \ \ \ \ \ \ \ \ \ \ \ \ \ \ \ \ \ \
\ \ \ \ \ \ \ \ \ \ \ \ \ \ \ \ \ \ \ \ \ \ \ \ \ \ \ \ \ \ \ \ \ \ \ \ \ \
\ \ \ \ \ \ \ \ \ \ \ \ \ \ \ \ \ \ \ \ \ \ \ \ \ \ \ \ \ \ \ \ \ \ \ \ \ \
\ \ \ \ \ \ \ \ \ \ \ \ \ \ \ \ \ \ \ \ \ \ \ \ \ \ \ \ \ \ \ \ \ \ \ \ \ \
\ \ \ \ \ \ \ \ \ \ \ \ \ \ \ \ \ \ \ \ \ \ \ \ \ \ \ \ \ \ \ \ \ \ \ \ \ \
\ \ \ \ \ \ \ \ \ \ \ \ \ \ \ \ \ \ \ \ \ \ \ \ \ \ \ \ \ \ \ \ \ \ \ \ \
\qquad\ \ \ \ \ \ \ \ \ \ \ \ \ \ \ \ \ \ \ \ \ \ \ \ \ \ \ \ \ \ \ \ \ \ \
\ \ \ \ \ \ \ \ \ \ \ \ \ \ \ \ \ \ \ \ \ \ \ \ \ \ \ \ \ \ \ \ \ \ \ \ \ \
\ \ \ \ \ \ \ \ \ \ \ \ \ \ \ \ \ \ \ \ \ \ \ \ \ \ \ \ \ \ \ \ \ \ \ \ \ \
\ \ \ \ \ \ \ \ \ \ \ \ \ \ \ \ \ \ \ \ \ \ \ \ \ \ \ \ \ \ \ \ \ \ \ \ \ \
\ \ \ \ \ \ \ \ \ \ \ \ \ \ \ \ \ \ \ \ \ \ \ \ \ \ \ \ \ \ \ \ \ \ \ \ \ \
\ \ \ \ \ \ \ \ \ \ \ \ \ \ \ \ \ \ \ \ \ \ \ \ \ \ \ \ \ \ \ \ \ \ \ \ \ \
\ \ \ \ \ \ \ \ \ \ \ \ \ \ \ \ \ \ \ \ \ \ \ \ \ \ \ \ \ \ \ \ \ \ \ \ \ \
\ \ \ \ \ \ \ \ \ \ \ \ \ \ \ \ \ \ \ \ \ \ \ \ \ \ \ \ \ \ \ \ \ \ \ \ \ \
\ \ \ \ \ \ \qquad\ \ \ \ \ \ \ \ \ \ \ \ \ \ \ \ \ \ \ \ \ \ \ \ \ \ \ \ \
\ \ \ \ \ \ \ \ \ \ \ \ \ \ \ \ \ \ \ \ \ \ \ \ \ \ \ \ \ \ \ \ \ \ \ \ \ \
\ \ \ \ \ \ \ \ \ \ \ \ \ \ \ \ \ \ \ \ \ \ \ \ \ \ \ \ \ \ \ \ \ \ \ \ \ \
\ \ \ \ \ \ \ \ \ \ \ \ \ \ \ \ \ \ \ \ \ \ \ \ \ \ \ \ \ \ \ \ \ \ \ \ \ \
\ \ \ \ \ \ \ \ \ \ \ \ \ \ \ \ \ \ \ \ \ \ \ \ \ \ \ \ \ \ \ \ \ \ \ \ \ \
\ \ \ \ \ \ \ \ \ \ \ \ \ \ \ \ \ \ \ \ \ \ \ \ \ \ \ \ \ \ \ \ \ \ \ \ \ \
\ \ \ \ \ \ \ \ \ \ \ \ \ \ \ \ \ \ \ \ \ \ \ \ \ \ \ \ \ \ \ \ \ \ \ \ \ \
\ \ \ \ \ \ \ \ \ \ \ \ \ \ \ \ \ \ \ \ \ \ \ \ \ \ \ \ \ \ \ \ \ \ \ \ \ \
\ \ \ \ \ \ \ \ \ \ \ \ \ \ \ \ \ \ \ \ \ \ \ \ \ \ \ \ \ \ \ \ \ \ \ \ \ \
\ \ \ \ \ \ \ \ \ \ \ \ \ \ \ \ \ \ \ \ \ \ \ \qquad\ \ \ \ \ \ \ \ \ \ \ \
\ \ \ \ \ \ \ \ \ \ \ \ \ \ \ \ \ \ \ \ \ \ \ \ \ \ \ \ \ \ \ \ \ \ \ \ \ \
\ \ \ \ \ \ \ \ \ \ \ \ \ \ \ \ \qquad

In order to deduce the\textbf{\ }short-lived quarks from energy bands, we
assume the following phenomenological formulae for the rest mass and
intrinsic quantum numbers (I, S, C, B and Q) of energy bands:

1). For a group of degenerate energy bands (number = deg) with the same
energy and equivalent $\overrightarrow{n}$ values (\ref{nnn}), the isospin is

\begin{equation}
\text{2I + 1 = }\deg \rightarrow \text{I = }\frac{\text{deg - 1}}{\text{2}}
\label{IsoSpin}
\end{equation}

2). The strange number S of an energy band (quark) that lies on an axis with
a rotary fold R of the regular rhombic dodecahedron is 
\begin{equation}
\text{S = R - 4.}  \label{S-Number}
\end{equation}

3). For energy bands with deg \TEXTsymbol{<} R and R - deg $\neq $ 2 (such
as the single energy bands on the $\Gamma $-H axis and the $\Gamma $-N
axis), the strange number is 
\begin{equation}
\text{S = S}_{axis}\text{+ }\Delta \text{S, \ }\Delta \text{S = }\delta 
\text{(}\widetilde{n}\text{) + [1-2}\delta \text{(S}_{axis}\text{)]Sign(}%
\widetilde{n}\text{)}  \label{S+DS}
\end{equation}%
where $\delta $($\widetilde{n}$)\ and $\delta $(S$_{axis}$) are Dirac
functions and S$_{axis}$ is the strange number (\ref{S-Number}) of the axis.
For an energy band with $\overrightarrow{n}$ = (n$_{1}$, n$_{2}$, n$_{3})$, $%
\widetilde{n}$ \ is defined as 
\begin{equation}
\widetilde{n}\text{ }\equiv \frac{\text{n}_{1}\text{+n}_{2}\text{+n}_{3}}{%
\left\vert \text{n}_{1}\right\vert \text{+}\left\vert \text{n}%
_{2}\right\vert \text{+}\left\vert \text{n}_{3}\right\vert }\text{. Sgn(}%
\widetilde{n}\text{) = }\left[ \text{%
\begin{tabular}{l}
+1 for $\widetilde{n}$ \TEXTsymbol{>} 0 \\ 
0 \ \ for $\widetilde{n}$ = 0 \\ 
-1 \ for $\widetilde{n}$ \TEXTsymbol{<} 0%
\end{tabular}
}\right]  \label{n/n}
\end{equation}

\begin{equation}
\text{If }\widetilde{n}\text{ = 0 \ \ \ \ }\Delta \text{S = }\delta \text{%
(0) = +1 \ from (\ref{S+DS}) and (\ref{n/n}).}  \label{n=0-DS=+1}
\end{equation}

\begin{equation}
\text{If \ }\widetilde{n}\text{ = }\frac{0}{0}\text{, we assume }\Delta 
\text{S = - S}_{Axis}\text{ .}  \label{DaltaS=-Sax}
\end{equation}%
Thus, for $\overrightarrow{n}$ = (0, 0, 0), from (\ref{DaltaS=-Sax}), we
have \ 
\begin{equation}
\text{S = S}_{Axis}\text{+ }\Delta \text{S = S}_{Axis}\text{- S}_{Axis}\text{
= 0.}  \label{S=0 of n=0}
\end{equation}%
\qquad \qquad \qquad\ \ \ \ 

4). If an energy band with S = +1, we call it has a charmed number C (C =
1): 
\begin{equation}
\text{if }\Delta \text{S = +1}\rightarrow \text{S = S}_{Ax}\text{+}\Delta 
\text{S = +1, }C\text{ }\equiv \text{+1.}  \label{Charmed}
\end{equation}%
If an energy band with S = -1$,$ which originates from $\Delta S=+1$ (S$%
_{Ax} $= -2), and there is an energy fluctuation,$\ $we call it has a bottom
number\ B:$\ \ \ \ \ \ \ \ \ \ \ \ \ $%
\begin{equation}
\text{for single bands, if\ }\Delta \text{S = +1}\rightarrow \text{S = -1
and }\Delta \text{E}\neq \text{0, B}\equiv \text{-1.}  \label{Battom}
\end{equation}

5). The elementary quark $\epsilon _{u}$ (or $\epsilon _{d}$) determines the
electric charge Q of an excited quark. For an excited quark of $\epsilon
_{u} $ (or $\epsilon _{d}$), Q = +$\frac{2}{3}$ (or -$\frac{1}{3}$). For an
excited quark with isospin I, there are 2I +1 members . For I$_{z}$ 
\TEXTsymbol{>} 0, Q = +$\frac{2}{3}$; I$_{z}$ \TEXTsymbol{<} 0, Q = -$\frac{1%
}{3}$ and I$_{z}$ = 0, 
\begin{eqnarray}
\text{ if S+C+b }\text{\TEXTsymbol{>} 0, Q = Q} &&_{\epsilon _{u}(0)}\text{
= }\frac{2}{3}\text{;}  \label{2/3} \\
\text{ if S+C+b }\text{\TEXTsymbol{<} 0, Q = Q} &&_{\epsilon _{d}(0)}\text{
= -}\frac{1}{3}\text{.}  \label{- 1/3}
\end{eqnarray}%
There is no quark with I$_{z}$ = 0 and S + C + b = 0.\ \ 

6). Since the experimental full width of baryons is about 100 Mev order, for
simplicity, we assume that a fluctuation energy $\Delta $E of a quark
roughly is

\begin{equation}
\Delta \text{E = 100 S[(1+S}_{Ax}\text{)(J}_{S,}\text{+S}_{Ax}\text{)]}%
\Delta \text{S \ \ \ J}_{S}\text{=}\left\vert \text{S}_{Ax}\right\vert +%
\text{1,2,3, ....}  \label{Dalta-E}
\end{equation}%
Fitting experimental results, we can get 
\begin{equation}
\alpha \text{ = 360 Mev.}  \label{360}
\end{equation}%
The rest mass (m$^{\ast }$) of a quark is the minimum energy of the energy
band. From (\ref{E(nk)}), (\ref{360}) and (\ref{Dalta-E}), the rest mass (m$%
^{\ast }$) \ of the quark is 
\begin{equation}
\begin{tabular}{l}
$\text{m}^{\ast }\text{ = \{313+ 360 minimum[(n}_{1}\text{-}\xi \text{)}^{2}%
\text{+(n}_{2}\text{-}\eta \text{)}^{2}\text{+(n}_{3}\text{-}\zeta \text{)}%
^{2}\text{]+}\Delta \text{E+}\Delta \text{\} (Mev)}$ \\ 
\ \ \ \ = m + $\Delta $ \ (Mev),%
\end{tabular}
\label{Rest Mass}
\end{equation}%
This formula (\ref{Rest Mass}) is the united quark mass formula.

Using above phenomenological formulae, we will show that each energy band
corresponds to a short-lived quark and deduce its intrinsic quantum numbers
I, S, C, B and Q.\textbf{\ }The minimum energy of the energy band is the
rest mass of the short-lived quark corresponding to the energy band.

\ \ \ \ \ \ \ \ \ \ \ \ \ \ \ \ \ \ \ \ \ \ \ \ \ \ 

\textbf{5 Deducing quarks using the phenomenological formulae from energy\
bands\ \ }

\ \ \ \ \ \ \ \ \ \ \ \ \ \ \ \ \ \ \ \ \ \ \ \ \ \ \ \ \ \ \ \ \ \ \ \ \ \
\ \ \ \ \qquad \qquad

\textbf{5.1 Deducing quarks from the energy bands in Tables 1 and 2 }

\ \ \ \ \ \ \ \ \ \ \qquad\ \ \ \ \ \ \ \ \ \ \ \ \ 

From deduced single energy bands of the $\Delta $-axis in Table 1, we can
use the above formulae (\ref{IsoSpin})-(\ref{Rest Mass})\ to deduce the
quarks. For the $\Delta $-axis, R = 4, strange number S$_{\Delta }$ = 0 from
(\ref{S-Number}). For single energy bands, I = 0 from (\ref{IsoSpin}); and S
= S$_{\Delta }$+ $\Delta $S = $\Delta $S = $\delta $($\widetilde{n}$) - Sign(%
$\widetilde{n}$) from (\ref{S+DS}). For $\overrightarrow{n}$ = (0, 0, -2)
and (0, 0, -4), $\Delta $S\ = +1 from (\ref{n/n}) and (\ref{S+DS}); for n =
(0, 0, 2), (0, 0, 4) and (0, 0, 6) $\Delta $S = -1 from (\ref{n/n}) and (\ref%
{S+DS}). Using (\ref{Charmed}), (\ref{n/n}) and (\ref{S+DS}), we can find
the charmed number C = +1 when n = (0, 0, -2) and (0, 0, -4). From (\ref{2/3}%
), we can find Q = $\frac{2}{3}$ when n = (0, 0, -2) and (0, 0, -4); from (%
\ref{- 1/3}), Q = - $\frac{1}{3}$ when n = (0, 0, 2), (0, 0, 4) and (0, 0,
6). From (\ref{Dalta-E}) and (\ref{Rest Mass}), we can find the rest masses
(minimE + $\Delta $E) . We list all results in Table 3:

\ \ \ \ \ \ \ \ \ \ \ \ \ \ \ \ \ \ \ \ 

\begin{tabular}{|l|}
\hline
\ \ \ \ \ \ \ \ \ \ Table 3. The u$_{C}$(m$^{\ast }$)-quarks and the d$_{S}$%
(m$^{\ast }$)-quarks on the $\Delta $-axis \\ \hline
S$_{axis}$ = 0, I = 0, S = $\Delta $S = $\delta $($\widetilde{n}$) + [1-2$%
\delta $(S$_{axis}$)]Sign($\widetilde{n}$), $\widetilde{n}\text{ }\equiv 
\frac{\text{n}_{1}\text{+n}_{2}\text{+n}_{3}}{\left\vert \text{n}%
_{1}\right\vert \text{+}\left\vert \text{n}_{2}\right\vert \text{+}%
\left\vert \text{n}_{3}\right\vert }$ \\ \hline
\begin{tabular}{|l|l|l|l|l|l|l|l|l|l|l|}
\hline
$\text{n}_{1,}\text{n}_{2,}\text{n}_{3}$ & $\text{E}_{Point}$ & MinimE & $%
\Delta \text{S}$ & J & I & S & C & Q & $\Delta \text{E}$ & $q_{\text{Name}%
}(m^{\ast })$ \\ \hline
$\text{{\small 0,\ \ 0, \ 0}}$ & $\text{E}_{\Gamma }\text{=0}$ & 313 & 0 & J$%
\text{ = 0}$ & $\frac{1}{2}$ & 0 & 0 & $\frac{2}{3}$ & 0 & $\text{u(313+}%
\Delta \text{)}$ \\ \hline
$\text{{\small 0, \ 0, \ 2}}$ & $\text{E}_{H}\text{=1}$ & 673 & -1 & J$_{%
\text{S,H}}\text{ =1}$ & 0 & -1 & 0 & -$\frac{1}{3}$ & 100 & $\text{d}_{S}%
\text{(773+}\Delta \text{)}$ \\ \hline
$\text{{\small 0, \ 0, -2}}$ & $\text{E}_{\Gamma }\text{=4}$ & 1753 & +1 & J$%
_{\text{C,}\Gamma }\text{=1}$ & 0 & 0 & 1 & $\frac{2}{3}$ & 0 & $\text{u}_{C}%
\text{(1753+}\Delta \text{)}$ \\ \hline
$\text{{\small 0, \ 0, \ 4}}$ & $\text{E}_{H}\text{=9}$ & 3553 & -1 & J$_{%
\text{S,H}}\text{ =2}$ & 0 & -1 & 0 & -$\frac{1}{3}$ & 200 & $\text{d}_{S}%
\text{(3753+}\Delta \text{)}$ \\ \hline
$\text{{\small 0, \ 0, -4}}$ & $\text{E}_{\Gamma }\text{=16}$ & 6073 & +1 & J%
$_{\text{C,}\Gamma }\text{=2}$ & 0 & 0 & 1 & $\frac{2}{3}$ & 0 & $\text{u}%
_{C}\text{(6073+}\Delta \text{)}$ \\ \hline
$\text{{\small 0, \ 0, \ 6}}$ & $\text{E}_{H}\text{=25}$ & $\text{9313}$ & -1
& J$_{\text{S,H}}\text{ =3}$ & 0 & -1 & 0 & -$\frac{1}{3}$ & 300 & $\text{d}%
_{S}\text{(9613+}\Delta \text{)}$ \\ \hline
\end{tabular}
\\ \hline
\end{tabular}

\ \ \ \ \ \ \ \ \ \ \ \ \ \ \ \ \ \ \ 

Similarly, for the $\Sigma $-axis, R = 2, strange number S$_{axis}$ = -2
from (\ref{S-Number}). For single energy bands, I = 0 from (\ref{IsoSpin}).
From (\ref{S+DS}), S = S$_{axis}$+ $\Delta $S = -2+ $\Delta $S; the $\Delta $%
S = $\delta $($\widetilde{n}$) + Sign($\widetilde{n}$). For $\overrightarrow{%
n}$ = (1, 1, 0), (2, 2, 0), (3, 3, 0) and (4, 4, 0), $\Delta $S\ = +1\ from (%
\ref{n/n}) and (\ref{S+DS}); for $\overrightarrow{n}$ =\ (-1, -1, 0), (-2,
-2, 0) and (-3, -3, 0), $\Delta $S\ = -1 from (\ref{n/n}) and (\ref{S+DS}).
Using (\ref{Battom}), (\ref{n/n}), (\ref{S+DS}) and (\ref{Dalta-E}), we can
find the bottom number B = -1 when $\overrightarrow{n}$ = (3, 3, 0) and (4,
4, 0). From (\ref{2/3}) and (\ref{- 1/3}),we can find the electric charge Q
= -$\frac{1}{3}$ for all quarks. From (\ref{Dalta-E}) and (\ref{Rest Mass}),
we can deduce rest masses (minimE + $\Delta $E) of quarks from the energy
bands in Table 2. We list all results in Table 4:

\ \ \ \ \ \ \ \ \ \ \ \ \ \ \ \ \ \ \ \ \ \ \ \ \ \ \ \ \ 

\begin{tabular}{|l|}
\hline
Table 4. The d$_{b}$(m$^{\ast }$)-Quarks, d$_{S}$(m$^{\ast }$)-Quarks and d$%
_{\Omega }$(m$^{\ast }$)-Quarks of the $\Sigma $-Axis \\ \hline
S$_{axis}$ = -2, I = 0, S = S$_{axis}+\Delta $S = $\delta $($\widetilde{n}$)
+ [1-2$\delta $(S$_{axis}$)]Sign($\widetilde{n}$), $\widetilde{n}\text{ }%
\equiv \frac{\text{n}_{1}\text{+n}_{2}\text{+n}_{3}}{\left\vert \text{n}%
_{1}\right\vert \text{+}\left\vert \text{n}_{2}\right\vert \text{+}%
\left\vert \text{n}_{3}\right\vert }$ \\ \hline
$%
\begin{tabular}{|l|l|l|l|l|l|l|l|l|l|l|}
\hline
E$_{Point}$ & $\text{\ n}_{1}\text{n}_{2}\text{n}_{3}$ & $\Delta \text{S}$ & 
S & B & Q & $\ \ \text{J}$ & I & minimE & $\Delta \text{E}$ & $\text{q}%
_{Name}\text{(m}^{\ast }\text{)}$ \\ \hline
$\text{E}_{\Gamma }\text{=0}$ & (0, 0, 0) & +2$^{\ast }$ & 0 & 0 & $\frac{-1%
}{3}$ & J$_{\text{S,}\Gamma }\text{ =0}$ & $\frac{1}{2}$ & 313 & 0 & d(313$%
\text{+}\Delta $) \\ \hline
$\text{E}_{N}\text{=}\frac{\text{1}}{2}$ & $(\text{1, 1, 0})$ & +1 & -1 & 0
& $\frac{-1}{3}$ & J$_{\text{S,N}}\text{ =1}$ & 0 & $\text{493}$ & 0 & $%
\text{d}_{S}\text{(493+}\Delta \text{)}$ \\ \hline
$\text{E}_{\Gamma }\text{=2}$ & $(\text{-1,-1,0})$ & - 1 & -3 & 0 & $\frac{-1%
}{3}$ & J$_{\text{S,}\Gamma }\text{ =1}$ & 0 & $\text{1033}$ & 0 & $\text{d}%
_{\Omega }\text{(1033+}\Delta \text{)}$ \\ \hline
$\text{E}_{N}\text{=}\frac{\text{9}}{2}$ & $(\text{2, 2, 0})$ & +1 & -1 & 0
& $\frac{-1}{3}$ & J$_{\text{S,N}}\text{ =2}$ & 0 & $\text{1933}$ & 0 & $%
\text{d}_{S}\text{(1933+}\Delta \text{)}$ \\ \hline
$\text{E}_{\Gamma }\text{=8}$ & $(\text{-2,-2,0})$ & - 1 & -3 & 0 & $\frac{-1%
}{3}$ & J$_{\text{S,}\Gamma }\text{ =2}$ & 0 & $\text{3193}$ & 0 & $\text{d}%
_{\Omega }\text{(3193+}\Delta \text{)}$ \\ \hline
$\text{E}_{N}\text{=}\frac{\text{25}}{2}$ & $(\text{3, 3, 0})$ & +1 & 0 & -1
& $\frac{-1}{3}$ & J$_{\text{S,N}}\text{ =3}$ & 0 & $\text{4813}$ & 100 & $%
\text{d}_{B}\text{(4913+}\Delta \text{)}$ \\ \hline
$\text{E}_{\Gamma }\text{=18}$ & $(\text{-3,-3,0})$ & - 1 & -3 & 0 & $\frac{%
-1}{3}$ & J$_{\text{S,}\Gamma }\text{ =3}$ & 0 & $\text{6793}$ & -300 & $%
\text{d}_{\Omega }\text{(6493+}\Delta \text{)}$ \\ \hline
$\text{E}_{N}\text{=}\frac{\text{49}}{2}$ & $(\text{4, 4, 0})$ & +1 & 0 & -1
& $\frac{-1}{3}$ & J$_{\text{S,N}}\text{ =4}$ & 0 & $\text{9133}$ & 200 & $%
\text{d}_{B}\text{(9333+}\Delta \text{)}$ \\ \hline
\end{tabular}%
\ $ \\ \hline
$^{\ast }$For $\overrightarrow{n}\text{ = (n}_{1}\text{, n}_{2}\text{, n}%
_{3})$ = (0, 0, 0), $\Delta $S = - S$_{axis}$= +2 from (\ref{DaltaS=-Sax})
\\ \hline
\end{tabular}

\ \ \ \ \ \ \ \ \ \ \ \ \ \ \ \ \ \ \ \ \ \ \ \ \ \ \ \ \ \ \ \ \ \ \ \ \ \
\ \ \ \ \ \ \ \ \ \ \ \ \ \ \ \ \ \ \ \ \ \ \ \ \ \ \ 

\textbf{5.2 The five deduced Ground quarks}

\ \ \ \ \ \ \ \ \ \ \ \ \ \ \ \ \ \ \ \ \ \ \ \ \ \ \ \ \ \ \ \ \ \ \ \ \ \
\ \ \ \ \ \ \ \ \ \ \ \ \ 

From Table 3 and Table 4, we find: The unflavored (S = C = B = 0) ground
quarks are u(313+$\Delta $) and d(313+$\Delta $). The strange quarks d$_{s}$%
(493), d$_{s}$(773), d$_{s}$(1933), d$_{S}$(3753) d$_{s}$(9613), $\text{d}%
_{\Omega }\text{(1033+}\Delta \text{), d}_{\Omega }\text{(3193+}\Delta \text{%
) and d}_{\Omega }\text{(6493+}\Delta \text{); The ground strange quark is }$%
d$_{s}$(493). The charmed quarks u$_{c}$(1753) and u$_{c}$(6073); the
charmed ground quark is u$_{c}$(1753). The bottom quarks d$_{b}$(4913) and d$%
_{b}$(9333); the bottom ground quark is d$_{b}$(4913). (in Table 11 of \cite%
{0502091} we have shown all low energy quarks, the five deduced ground
quarks are still the ground quarks of all quarks). For the four flavored
quarks, there are five ground quarks [u(313+$\Delta $), d(313+$\Delta $), d$%
_{S}$(493+$\Delta $), u$_{C}$(1753+$\Delta $) and d$_{b}$(4913+$\Delta $)]
in the quark spectrum. These five ground quarks correspond to the five
quarks \cite{Quarks} of the current Quark Model: u $\leftrightarrow $ u(313+$%
\Delta $), d $\leftrightarrow $ d(313+$\Delta $), s $\leftrightarrow $ d$%
_{S} $(493+$\Delta $), c $\leftrightarrow $ u$_{C}$(1753+$\Delta $) and b $%
\leftrightarrow $ d$_{b}$(4913+$\Delta $).

We can compare the rest masses and intrinsic quantum numbers (I, S, C, b and
Q) \cite{Quarks} of the current quarks with the deduced values of the five
ground quarks. The deduced intrinsic quantum numbers (I, S, C, b and Q ) of
the five ground quarks are exactly the same as the five current quarks as
shown in Table 5A:

\ \ \ \ \ \ \ \ \ \ \ \ \ \ \ \ \ \ \ \ 

\begin{tabular}{l}
\ \ \ \ \ \ \ Table 5A. The Five Deduced Ground Quarks and Current Quarks \\ 
$%
\begin{tabular}{|l|l|l|l|l|l|}
\hline
Dq(m)$^{\$}$,Cq$^{\ast }$ & u(313), u & d(313), d & d$_{s}$(493), s & u$_{c}$%
(1753), c & d$_{b}$(4913), b \\ \hline
Strange S & 0 \ \ \ \ \ \ \ \ \ 0 & 0 \ \ \ \ \ \ \ \ \ 0 & -1 \ \ \ \ \ \ \
-1 & 0 \ \ \ \ \ \ \ \ \ \ \ \ 0 & 0 \ \ \ \ \ \ \ \ \ \ \ \ \ 0 \\ \hline
Charmed C & 0 \ \ \ \ \ \ \ \ \ 0 & 0 \ \ \ \ \ \ \ \ \ 0 & 0 \ \ \ \ \ \ \
\ \ 0 & 1 \ \ \ \ \ \ \ \ \ \ \ \ 1 & 0 \ \ \ \ \ \ \ \ \ \ \ \ \ 0 \\ \hline
Bottom B & 0 \ \ \ \ \ \ \ \ \ 0 & 0 \ \ \ \ \ \ \ \ \ 0 & 0 \ \ \ \ \ \ \ \
\ 0 & 0 \ \ \ \ \ \ \ \ \ \ \ \ 0 & -1 \ \ \ \ \ \ \ \ \ \ \ \ -1 \\ \hline
Isospin I & $\frac{1}{2}$ \ \ \ \ \ \ \ \ $\frac{1}{2}$ & $\frac{1}{2}$ \ \
\ \ \ \ \ \ $\frac{1}{2}$ & 0 \ \ \ \ \ \ \ \ \ 0 & 0 \ \ \ \ \ \ \ \ \ \ \
\ 0 & 0 \ \ \ \ \ \ \ \ \ \ \ \ \ 0\  \\ \hline
I$_{Z}$ & $\frac{1}{2}$ \ \ \ \ \ \ \ \ $\frac{1}{2}$ & -$\frac{1}{2}$ \ \ \
\ \ -$\frac{1}{2}$ & 0 \ \ \ \ \ \ \ \ \ 0 & $0$ \ \ \ \ \ \ \ \ \ \ \ \ 0\ 
& 0 \ \ \ \ \ \ \ \ \ \ \ \ \ 0 \\ \hline
Electric $\text{Q}_{q}$ & $\frac{2}{3}$ \ \ \ \ \ \ \ \ $\frac{2}{3}$ & -$%
\frac{1}{3}$ \ \ \ \ \ -$\frac{1}{3}$ & -$\frac{1}{3}$ \ \ \ \ \ \ -$\frac{1%
}{3}$\  & $\frac{2}{3}$ \ \ \ \ \ \ \ \ \ \ \ $\frac{2}{3}$ & -$\frac{1}{3}$
\ \ \ \ \ \ \ \ \ \ -$\frac{1}{3}$\  \\ \hline
\end{tabular}%
\ $ \\ 
\ \ \ Dq(m)$^{\$}$= Deduced ground quark; Cq$^{\ast }$= Current quark%
\end{tabular}

\ \ \ \ \ \ \ \ \ \ \ \ \ \ \ \ \ \ \ \ \ \ \ \ \ \ 

The deduced rest masses of the five ground quarks are roughly a constant
(about 390 Mev) larger than the masses of the current quarks, as shown in
Table 5B.

\ \ \ \ \ \ \ \ \ \ \ \ \ \ \ \ \ \ \ \ \ \ \ \ \ \ \ \ \ \ \ \ \ \ \ `\ \ \
\ \ \ \ \ \ \ \ \ \ \ \ \ \ \ \ \ \ 

\begin{tabular}{l}
\ Table 5B Comparing the Rest Masses of Deduced and Current Quarks \\ 
$%
\begin{tabular}{|l|l|l|l|l|l|}
\hline
Current Quark & Up & Down & Strange & Charmed & bottom \\ \hline
Current Quark(m) & u(2.8) & d(6) & s(105) & c(1225) & b(4500) \\ \hline
Current quark mass & {\small 1.5 to 4} & {\small 4 to 8} & {\small 80 to 130}
& {\small 1250 to 1350} & 
\begin{tabular}{l}
{\small 4.1 to 4.4 G.} \\ 
{\small 4.6 to 4.9 G.}%
\end{tabular}
\\ \hline
Deduced Quark (m) & u(313) & d(313) & d$_{S}$(493) & u(1753) & d$_{b}$(4913)
\\ \hline
$\left\vert \text{m}_{Curr.}\text{-m}_{Deduced.}\right\vert $ & 310 & 307 & 
388 & 528 & 413 \\ \hline
\end{tabular}%
\ $ \\ 
The rest mass of a deduced quark m$^{\ast }$= m + $\Delta \ \rightarrow $ m
= m$^{\ast }$ - $\Delta $%
\end{tabular}

\ \ \ \ \ \ \ \ \ \ \ \ \ \ \ \ \ \ \ \ \ \ \ \ \ \ \ \ \ \ 

These mass differences may originate from different energy reference
systems. If we use the same energy reference system, the deduced masses of
ground quarks will be roughly consistent with the masses of the
corresponding current quarks. Of course, the ultimate test is whether or not
the baryons and mesons composed of the deduced quarks are consistent with
experimental results.\ \ \ \ \ \ \ \ \ \ \ \ \ \ \ \ \ 

We will deduce the baryons and mesons composed of the quarks in Table 3 and
4 in this paper.

\ \ \ \ \ \ \ \ \ \ \ \ \ \ \ \ \ \ \ \ \ \ \ \ \ \ \ \ \ \ \ \ \ \ \ \ 

\textbf{6 Deducing the baryons of the quarks in Table 3 and 4}

\ \ \ \ \ \ \ \ \ \ \ \ \ \ \ \ \ \ \ \ \ \ \ \ \ \ \ \ \ \ \ \ \ \ \ \ 

According to the Quark Model \cite{Quark Model}, a colorless baryon is
composed of three different\ colored quarks. Using sum laws (\ref{SumB}) and
the deduced quarks in Tables 3 and 4, we can deduce the baryons as shown in
Table 6. From Tables 3 and 4, we can see that there is a term $\Delta \ $of
the rest masses of quarks. $\Delta $ is a very large unknown constant. Since
the rest masses of the quarks in a baryon are large (from $\Delta $) and the
rest mass of the baryon composed by three quarks is not, we infer that there
will be a strong binding energy (E$_{Bind}$ = - 3$\Delta $) to cancel 3$%
\Delta $ from the three quarks: M$_{\text{B}}\text{ = m}_{q_{1}}^{\ast }$ +
\ m$_{q_{_{N}(313)}}^{\ast }$ \ + m$_{_{q_{_{N}(313)}}}^{\ast }$- $%
\left\vert \text{E}_{Bind}\right\vert $ = m$_{q_{1}}$+ m$_{q_{_{N}(313)}}$+ m%
$_{_{q_{_{N}(313)}}}$+ 3$\Delta $ - 3$\Delta $ = m$_{q_{1}}$+ m$%
_{q_{_{N}(313)}}$+ m$_{_{q_{_{N}(313)}}}$. Thus we will omit the term 3$%
\Delta \ $in the three quark masses and the term $-3\Delta \ $in the binding
energy from now on.\ For simplicity's sake, we only deduced baryons composed
of at least two free excited quark q$_{N}$(313) [u(313), d(313)] since other
baryons have much lower productivity. For these baryons, sum laws are:

\begin{equation}
\begin{tabular}{l}
Baryon strange number $\ \text{S}_{\text{B}}\text{ }\text{= S}_{q_{1}}\text{%
+ S}_{q_{_{N}(313)}}\text{+ S}_{q_{_{N}(313)}}=\text{ S}_{q_{1}}\text{,}$ \\ 
baryon charmed number $\text{C}_{\text{B\ }}\text{ }\text{= C}_{q_{1}}\text{
+ C}_{q_{_{N}(313)}}\text{ + C}_{q_{_{N}(313)}}\text{= C}_{q_{1}}\text{,}$
\\ 
baryon bottom number $\ $B$_{\text{B}}\text{=}\text{ B}_{q_{1}}\text{+ B}%
_{q_{_{N}(313)}}\text{+ B}_{q_{_{N}(313)}}\text{= B}_{q_{1}}\text{,}$ \\ 
baryon electric charge $\ \text{Q}_{\text{B}}\text{ = Q}_{q_{1}}\text{+ Q}%
_{q_{_{N}(313)}}\text{+ Q}_{_{q_{_{N}(313)}}}.$ \\ 
baryon mass M$_{\text{B}}$ = \ $\text{m}_{q_{1}}\text{+ m}_{q_{_{N}(313)}}%
\text{+ m}_{_{q_{_{N}(313)}}}($except charmed baryons)%
\end{tabular}
\label{SumB}
\end{equation}%
For charmed baryons, M$_{\text{B}}$ = \ $\text{m}_{q_{1}}\text{+ m}%
_{q_{_{N}(313)}}\text{+ m}_{_{q_{_{N}(313)}}}$+ $\Delta e$%
\begin{equation}
\Delta e\text{ = 100C(2I-1)}  \label{Ebin of B}
\end{equation}%
where C is the charmed number and I is the isospin of the charmed baryons.
Using sum laws (\ref{SumB}) and binding energy formula (\ref{Ebin of B}), we
deduce the baryons shown in Table 6 from the quarks in Tables 3 and 4:

\ \ \ \ \ \ \ \ \ \ \ \ \ \ \ \ \ \ \ \ \ \ \ \ \ \ \ \ \ 

\begin{tabular}{l}
\ \ \ \ \ \ \ \ \ \ Table 6. The Baryons of the Quarks in Table 3 and Table 4
\\ 
\begin{tabular}{|l|l|l|l|l|l|l|l|l|l|l|l|}
\hline
{\small q}$_{i}^{I_{z}}$(m) & {\small q}$_{j}$ & {\small q}$_{k}$ & {\small I%
} & {\small S} & {\small C} & {\small b} & {\small Q} & {\small M} & {\small %
Baryon} & {\small Exper.} & $\frac{\Delta M}{M}${\small \%} \\ \hline
{\small u}$^{\frac{1}{2}}${\small (313)} & {\small u} & {\small d} & $\frac{1%
}{2}$ & {\small 0} & {\small 0} & {\small 0} & {\small 1} & {\small 939} & 
{\small p(939)} & {\small p(938)} & {\small 0.11} \\ \hline
{\small d}$^{\frac{-1}{2}}${\small (313)} & {\small u} & {\small d} & $\frac{%
1}{2}$ & {\small 0} & {\small 0} & {\small 0} & {\small 0} & {\small 939} & 
{\small n(939)} & {\small n(940)} & {\small 0.11} \\ \hline
{\small d}$_{s}^{0}${\small (493)} & {\small u} & {\small d} & {\small 0} & 
{\small -1} & {\small 0} & {\small 0} & {\small 0} & {\small 1119} & $%
\Lambda ${\small (1119)} & $\Lambda ^{0}${\small (1116)} & {\small 0.27} \\ 
\hline
{\small u}$_{c}^{0}${\small (1753)} & {\small u} & {\small d} & {\small 0} & 
{\small 0} & {\small 1} & {\small 0} & {\small 1} & {\small 2279} & $\Lambda
_{c}${\small (2279)} & $\Lambda _{c}^{+}${\small (2285)} & {\small 0.3.} \\ 
\hline
{\small u}$_{c}^{0}${\small (1753)} & {\small u} & {\small u} & {\small 1} & 
{\small 0} & {\small 1} & {\small 0} & {\small 2} & {\small 2479} & $\Sigma
_{c}^{++}${\small (2479)} & $\Sigma _{c}^{++}${\small (2455)} & {\small 1.0}
\\ \hline
{\small u}$_{c}^{0}${\small (1753)} & {\small u} & {\small d} & {\small 1} & 
{\small 0} & {\small 1} & {\small 0} & {\small 1} & {\small 2479} & $\Sigma
_{c}^{+}${\small (2479)} & $\Sigma _{c}^{+}${\small (2455)} & {\small 1.0}
\\ \hline
{\small u}$_{c}^{0}${\small (1753)} & {\small d} & {\small d} & {\small 1} & 
{\small 0} & {\small 1} & {\small 0} & {\small 0} & {\small 2479} & $\Sigma
_{c}^{-}${\small (2479)} & $\Sigma _{c}^{-}${\small (2455)} & {\small 1.0}
\\ \hline
{\small d}$_{b}^{0}${\small (4913)} & {\small u} & {\small d} & {\small 0} & 
{\small 0} & {\small 0} & {\small -1} & {\small 0} & {\small 5539} & $%
\Lambda _{b}${\small (5539)} & $\Lambda _{b}^{0}${\small (5624)} & {\small %
1.5} \\ \hline
{\small d}$_{S}^{0}${\small (773)} & {\small u} & {\small d} & {\small 0} & 
{\small -1} & {\small 0} & {\small 0} & {\small 0} & {\small 1399} & $%
\Lambda ${\small (1399)} & $\Lambda ${\small (1405)} & {\small 0.4} \\ \hline
{\small d}$_{S}^{0}${\small (1933)} & {\small u} & {\small d} & {\small 0} & 
{\small -1} & {\small 0} & {\small 0} & {\small 0} & {\small 2559} & $%
\Lambda ${\small (2559)} & $\Lambda ${\small (2585)}$^{\ast \ast }$ & 
{\small 1.0} \\ \hline
{\small d}$_{S}^{0}${\small (3753)} & {\small u} & {\small d} & {\small 0} & 
{\small -1} & {\small 0} & {\small 0} & {\small 0} & {\small 4375} & $%
\Lambda ${\small (4375)} & {\small Prediction} &  \\ \hline
{\small d}$_{S}^{0}${\small (9613)} & {\small u} & {\small d} & {\small 0} & 
{\small -1} & {\small 0} & {\small 0} & {\small 0} & {\small 10239} & $%
\Lambda ${\small (10239)} & {\small Prediction} &  \\ \hline
{\small u}$_{c}^{0}${\small (6073)} & {\small u} & {\small d} & {\small 0} & 
{\small 0} & {\small 1} & {\small 0} & {\small 1} & {\small 6699} & $\Lambda
_{C}^{+}${\small (6699)} & {\small Prediction} &  \\ \hline
{\small d}$_{b}^{0}${\small (9333)} & {\small u} & {\small d} & {\small 0} & 
{\small 0} & {\small 0} & {\small -1} &  & {\small 9959} & $\Lambda _{b}^{0}$%
{\small (9959)} & {\small Prediction} &  \\ \hline
{\small d}$_{\Omega }^{0}${\small (1033)} & {\small d} & {\small d} & 
{\small 0} & {\small -3} & {\small 0} & {\small 0} & {\small -1} & {\small %
1659} & $\Omega ^{-}${\small (1659)} & $\Omega ^{-}${\small (1672)} & 
{\small 0.8} \\ \hline
\end{tabular}
\\ 
{\small u }$\equiv ${\small \ u(313) and d }$\equiv ${\small \ d(313). }$%
\Lambda ${\small (2585)}$^{\ast \ast }${\small \ Evidence of existence only
fair.}%
\end{tabular}

\ \ \ \ \ \ \ \ \ \ \ \ \ \ \ \ \ \ \ \ \ \ \ \ \ \ \ \ \ \ 

Table 6 shows that the deduced intrinsic quantum numbers ( I, S, C, b and Q)
of the baryons match experimental results \cite{Baryon04} exactly and that
the deduced rest masses of the baryons are consistent with more than 98.5\%
of experimental results \cite{Baryon04}. These are strong supports for the
deduced the rest masses and intrinsic quantum numbers of the quarks.

\ \ \ \ \ \ \ \ \ \ \ \ \ \ \ \ \ \ \ \ \ \ \ \ \ \ \ \ \ \ \ \ \ \ \ \ \ \
\ 

\textbf{7 Deducing the mesons of the quarks in Tables 3 and 4}

\ \ \ \ \ \ \ \ \ \ \ \ \ \ \ \ \ \ \ \ \ \ \ \ \ \ \ \ \ \ 

According to the Quark Model \cite{Quark Model}, a colorless meson is
composed of a quark q$_{i}$ with a color and an antiquark $\overline{q_{j}}$
with the anticolor of the quark q$_{i}$. For each flavor, the three
different colored quarks have the same I, S, C, B, Q and m. Thus we can omit
the color when we deduce the rest masses and intrinsic quantum numbers of
the mesons. For mesons, the sum laws are

\begin{equation}
\begin{tabular}{l}
Meson strange number $\text{S}_{\text{M}}\text{ }\text{= S}_{q_{i}}\text{+ S}%
_{\overline{q_{j}}}\text{,}$ \\ 
meson charmed numbers $\text{C}_{\text{M\ }}\text{ }\text{= C}_{q_{i}}\text{
+ C}_{\overline{q_{j}}}\text{ ,}$ \\ 
meson bottom number B$_{\text{M}}\text{=}\text{ B}_{q_{i}}\text{+ B}_{%
\overline{q_{j}}}\text{,}$ \\ 
meson electric charge $\text{Q}_{\text{M}}\text{ = Q}_{q_{i}}\text{+ Q}_{%
\overline{q_{j}}}.$%
\end{tabular}
\label{Sum(SCbQ)}
\end{equation}%
There is a strong interaction between the quark and antiquark (colors), but
we do not know how large it is. Since the rest masses of the quarks in
mesons are large (from $\Delta $) and the rest mass of the meson composed of
the quark and antiquark is not, we infer that there will be a large portion
of binding energy (- 2$\Delta $) to cancel 2$\Delta $ from the quark and
antiquark and a small amount of binding energy as shown in the following 
\begin{equation}
\text{E}_{B}\text{(q}_{i}\overline{q_{j}}\text{) = -2}\Delta \text{ - 337 +
100[}\frac{\Delta \text{m}}{\text{m}_{g}}\text{ +DS -}\ \widetilde{m}\text{+}%
\gamma \text{(i,j) -2I}_{i}\text{I}_{j}\text{]}  \label{Eb-Meson}
\end{equation}%
where $\Delta $ = $\frac{1}{3}\left\vert \text{E}_{bind}\right\vert $ (\ref%
{Dalta}) is $\frac{1}{3}$ of the binding energy of a baryon (an unknown
large constant, $\Delta $ \TEXTsymbol{>}\TEXTsymbol{>} m$_{\text{P}}$= 938
Mev), $\Delta $m = $\left\vert \text{m}_{i}\text{-m}_{j}\right\vert $, DS =$%
\left\vert \text{(}\Delta \text{S)}_{i}\text{- (}\Delta \text{S)}%
_{j}\right\vert $. m$_{g}$ = 939 (Mev) unless 
\begin{equation}
\begin{tabular}{|l|l|l|l|}
\hline
m$_{i}$(or m$_{j}$) equals & m$_{C}\geqslant $ 6073 & m$_{b}\geqslant $ 9333
& m$_{S}\geqslant $ 9613 \\ \hline
$\ \ \ \text{m}_{g}${\small \ will equal to} & 1753(Table 4) & 4913 (Table7)
& 3753(Table 4). \\ \hline
\end{tabular}
\label{m(g)}
\end{equation}

$\ \widetilde{m}$ = $\frac{m_{i}\times m_{j}}{\text{m}_{g_{i}}\times \text{m}%
_{g_{j}}}$ \ \ m$_{g_{i}}$ = m$_{g_{j}}^{\text{ \ }}$ = 939 (Mev) unless 
\begin{equation}
\begin{tabular}{|l|l|l|l|l|l|}
\hline
\ m$_{i}$(or m$_{j}$) & m$_{q_{_{N}}}$=313 & m$_{d_{s}}$=493 & m$%
_{u_{c}}\succeq $1753 & m$_{d_{S}}$\TEXTsymbol{>} 3753, & m$_{d_{b}}\succeq $
4913 \\ \hline
\ m$_{g_{j}}$ (or m$_{g_{j}}^{\text{ \ }})$ & 313 & 493 & 1753 & 3753, & 
4913. \\ \hline
\end{tabular}
\label{M(gi)}
\end{equation}%
\ If\ q$_{i\text{ }}$and q$_{j}$ are both ground quarks in Table 5, $\gamma $%
(i, j) = 0. If\ q$_{i\text{ }}$and q$_{j}$ are not both ground quarks, for q$%
_{i\text{ }}$= q$_{j}$, $\gamma $(i, j) = -$1$; for q$_{i\text{ }}\neq $ \ q$%
_{j}$,$\ \gamma $(i, j) = +1. S$_{i}$ (or S$_{j}$) is the strange number of
the quark q$_{i}$ (or q$_{j}$). I$_{i}$ (or I$_{j}$) is the isospin of the
quark q$_{i}$ (or q$_{j}$).

From the quarks in Tables 3 and 4, we can use (\ref{Sum(SCbQ)}) and (\ref%
{Eb-Meson}) to deduce the rest masses and the intrinsic\ quantum numbers (I,
S, C, b and Q) of mesons as shown in Table 7.

\begin{tabular}{l}
$\ \ \ \ \ \ \ \ \ \ \ \text{Table 7.\ \ The Deduced Mesons of the Quarks in
Tables 3 and 4}$ \\ 
$%
\begin{tabular}{|l|l|l|l|l|l|l|l|}
\hline
$\text{q}_{i}^{\Delta S}\text{(m}_{i}\text{)}{\small \ }\overline{\text{q}%
_{j}^{\Delta S}\text{(m}_{j}\text{)}}$ & $\frac{100\Delta \text{m}}{\text{939%
}}$ & {\small DS} & {\small -100}$\widetilde{m}$ & {\small E}$_{bind}$ & 
{\small Deduced} & {\small Experiment} & {\small R} \\ \hline
$\text{q}_{N}^{0}${\small (313)}$\overline{_{N}^{0}\text{{\small (313)}}}$ & 
{\small 0} & ${\small 0}$ & {\small -100} & {\small - 487}$^{\#}$ & $\pi $%
{\small (139)} & $\pi ${\small (138)} & {\small 0.7} \\ \hline
{\small q}$_{N}^{0}${\small (313)}$\overline{\text{d}_{S}^{1}\text{(493)}}$
& {\small 19} & {\small 1} & {\small -100} & {\small - 318} & {\small K(488)}
& {\small K(494)} & {\small 0.2} \\ \hline
{\small d}$_{S}^{1}${\small (493)}$\overline{\text{d}_{S}^{1}\text{(493)}}$
& {\small 0} & {\small 0} & {\small -100} & {\small - 437} & $\eta ${\small %
(549)} & $\eta ${\small (548)} & {\small 0.2} \\ \hline
{\small u}$_{C}^{1}${\small (1753)}$\overline{\text{q}_{N}^{0}\text{{\small %
(313)}}}$ & {\small 153} & {\small 1} & {\small -100} & {\small - 184} & 
{\small D(1882)} & {\small D(1869)} & {\small 0.7} \\ \hline
{\small u}$_{C}^{1}${\small (1753)}$\overline{\text{q}_{S}^{1}\text{{\small %
(493)}}}$ & {\small 134} & {\small 0} & {\small -100} & {\small - 303} & 
{\small D}$_{S}${\small (1943)} & {\small D}$_{S}${\small (1969)} & {\small %
0.4} \\ \hline
{\small u}$_{C}^{1}${\small (1753)}$\overline{\text{u}_{C}^{1}\text{({\small %
1753})}}$ & {\small 0} & {\small 0} & {\small -100} & {\small -437} & 
{\small J/}$\psi ${\small (3069)} & {\small J/}$\psi ${\small (3097)} & 
{\small 0.9} \\ \hline
$\text{q}_{N}^{0}${\small (313)}$\overline{\text{d}_{b}^{1}\text{{\small %
(4913)}}}$ & {\small 490} & {\small 1} & {\small -100} & {\small \ 153}$%
^{\ast }$ & {\small B(5379)} & {\small B(5279)} & {\small 1.9} \\ \hline
{\small d}$_{S}^{1}${\small (493)}$\overline{\text{d}_{b}^{1}\text{{\small %
(4913)}}}$ & {\small 471} & {\small 0} & {\small -100} & {\small \ 34}$%
^{\ast }$ & {\small B}$_{S}${\small (5440)} & {\small B}$_{S}${\small (5370)}
& {\small 1.3} \\ \hline
{\small u}$_{C}^{1}${\small (1753)}$\overline{\text{d}_{b}^{1}\text{{\small %
(4913)}}}$ & {\small 337} & {\small 0} & {\small -100} & {\small -100} & 
{\small B}$_{C}${\small (6566)} & {\small B}$_{C}${\small (6400)} & {\small %
2.6} \\ \hline
{\small d}$_{b}^{1}${\small (4913)}$\overline{\text{d}_{b}^{1}\text{{\small %
(4913)}}}$ & {\small 0} & {\small 0} & {\small -100} & {\small -\ 437} & $%
\Upsilon ${\small (9389)} & $\Upsilon ${\small (9460)} & {\small 0.8} \\ 
\hline
{\small d}$_{S}^{-1}${\small (773)}$\overline{\text{d}_{S}^{-1}\text{{\small %
(773)}}}$ & {\small 0} & {\small 0} & {\small -68} & {\small -505} & $\eta $%
{\small (1041)} & $\phi ${\small (1020)} & {\small 2.0} \\ \hline
{\small d}$_{S}^{-1}${\small (3753)}$\overline{\text{d}_{S}^{-1}\text{%
{\small (3753)}}}$ & {\small 0} & {\small 0} & {\small -1597} & {\small -2034%
} & $\eta ${\small (5472)} & {\small prediction} &  \\ \hline
{\small d}$_{S}^{1}${\small (1933)}$\overline{\text{d}_{S}^{1}\text{{\small %
(1933)}}}$ & {\small 0} & {\small 0} & {\small -424} & {\small -861} & $\eta 
${\small (3005)} & $\eta _{c}${\small (2980)} & {\small 0.8} \\ \hline
{\small d}$_{S}^{1}${\small (9333)}$\overline{\text{d}_{S}^{1}\text{{\small %
(9333)}}}$ & {\small 0} & {\small 0} & {\small -361} & {\small -798} & $%
\Upsilon ${\small (17868)} & {\small prediction} & {\small ?} \\ \hline
{\small u}$_{C}^{1}${\small (6073)}$\overline{\text{u}_{C}^{1}\text{{\small %
(6073)}}}$ & {\small 0} & {\small 0} & {\small -1200} & {\small -1637} & $%
\psi ${\small (10509)} & $\Upsilon ${\small (10355)} & {\small 1.5} \\ \hline
{\small d}$_{S}^{-1}${\small (9613)}$\overline{\text{d}_{S}^{-1}\text{%
{\small (9613)}}}$ & {\small 0} & {\small 0} & {\small -656} & {\small -1093}
& $\eta ${\small (18133)} & {\small prediction} &  \\ \hline
{\small d}$_{\Omega }^{-1}${\small (1033)}$\overline{\text{d}_{\Omega }^{-1}%
\text{{\small (1033)}}}$ & {\small 0} & {\small 0} & {\small -121} & {\small %
-558} & $\eta ${\small (1508)} & {\small f}$_{0}${\small (1507)} & {\small %
0.7} \\ \hline
{\small q}$_{N}^{0}${\small (313)}$\overline{\text{d}_{S}^{-1}\text{{\small %
(773)}}}$ & {\small 49} & {\small 1} & {\small -82} & {\small -170} & 
{\small K(916)} & {\small K(892)} & {\small 2.7} \\ \hline
{\small q}$_{N}^{0}${\small (313)}$\overline{\text{d}_{S}^{1}\text{{\small %
(1933)}}}$ & {\small 171} & {\small 1} & {\small -206} & {\small -170} & 
{\small K(2076)} & {\small K}$_{4}^{\ast }${\small (2045)} & {\small 1.5} \\ 
\hline
{\small q}$_{N}^{0}${\small (313)}$\overline{\text{d}_{S}^{-1}\text{(3753)}}$
& {\small 347} & {\small 1} & {\small -400} & {\small -190} & {\small K(3876)%
} & {\small prediction} & {\small ?} \\ \hline
{\small q}$_{N}^{0}${\small (313)}$\overline{\text{d}_{S}^{-1}\text{{\small %
(9613)}}}$ & {\small 248} & {\small 1} & {\small -256} & {\small -145} & 
{\small K(9781)} & {\small prediction} & {\small ?} \\ \hline
{\small q}$_{N}^{0}${\small (313)}$\overline{\text{d}_{b}^{1}\text{{\small %
(9333)}}}$ & {\small 183} & {\small 1} & {\small -190} & {\small -144} & 
{\small B(9502)} & {\small prediction} & {\small ?} \\ \hline
{\small u}$_{C}^{1}${\small (6073)}$\overline{\text{q}_{N}^{0}\text{{\small %
(313)}}}$ & {\small 328} & {\small 1} & {\small -346.4} & {\small -155} & 
{\small D(6231)} & {\small prediction} & {\small ?} \\ \hline
{\small u}$_{C}^{1}${\small (6073)}$\overline{\text{d}_{S}^{1}\text{{\small %
(493)}}}$ & {\small 318.3} & {\small 0} & {\small -346.4} & {\small -265} & 
{\small D}$_{s}${\small (6301)} & {\small prediction} & {\small ?} \\ \hline
{\small d}$_{S}^{1}${\small (493)}$\overline{\text{d}_{S}^{-1}\text{{\small %
(773)}}}$ & {\small 30} & {\small 2} & {\small -256.1} & {\small -90} & $%
\eta ${\small (1177)} & $\eta ${\small (1170)} & {\small 0.6} \\ \hline
{\small d}$_{S}^{1}${\small (493)}$\overline{\text{d}_{S}^{-1}\text{{\small %
(3753)}}}$ & {\small 347} & {\small 2} & {\small -339.7} & {\small -90} & $%
\eta ${\small (4156)} & $\psi ${\small (4159)} & {\small .07} \\ \hline
{\small d}$_{S}^{1}${\small (493)}$\overline{\text{d}_{S}^{-1}\text{{\small %
(9613)}}}$ & {\small 243} & {\small 2} & {\small -256.1} & {\small -50} & $%
\eta ${\small (10056)} & $\Upsilon ${\small (10023)} & {\small 0.4} \\ \hline
{\small u}$_{C}^{1}${\small (1753)}$\overline{\text{d}_{S}^{-1}\text{{\small %
(773)}}}$ & {\small 104} & {\small 2} & {\small -82.3} & {\small -15} & 
{\small D}$_{S}${\small (2511)} & {\small D}$_{S_{1}}${\small (2535)} & 
{\small 1.0} \\ \hline
{\small d}$_{S}^{1}${\small (9613)}$\overline{\text{d}_{S}^{-1}\text{{\small %
(773)}}}$ & {\small 235} & {\small 0} & {\small -211} & {\small -212} & $%
\eta ${\small (10174)} & $\chi ${\small (10232)} & {\small 0.6} \\ \hline
\end{tabular}%
$ \\ 
$^{\#}${\small For q}$_{N}^{0}${\small (313)}$\overline{_{N}^{0}\text{%
{\small (313)}}}${\small , I}$_{i}${\small I}$_{j}$ {\small = }$\frac{1}{4}%
\rightarrow $ {\small 100( -2I}$_{i}${\small I}$_{j}${\small ) = - 50} \\ 
$^{\ast }${\small The total binding energy (153-2}$\Delta ${\small ) and
(34-2}$\Delta ${\small ) are negative from (\ref{Eb-Meson})}%
\end{tabular}

\ \ \ \ \ \ \ \ \ \ \ \ \ \ \ \ \ \ \ \ \ \ \ \ \ \ \ \ \ \ \ \ \ \ \ \ \ \
\ \ \ \ \ 

Table 7 shows that the deduced intrinsic quantum numbers match experimental
results \cite{Meson04} exactly. The deduced rest masses are consistent with
experimental results.

\ \ \ \ \ \ \ \ \ \ \ \ \ \ \ \ \ \ 

\textbf{8 Predictions\ }{\LARGE \ \ }\ \ \ \ \ \ \ \ \ \ \ \ \ \ \ \ \ \ \ \
\ \ \ \ \ \ \ \ \ \ \ \ \ \ \ \ \ \ \ \ \ \ \ \ \ \ \ \ \ \ \ \ \ \ 

\ \ \ \ \ \ \ \ \ \ \ \ \ \ \ \ \ \ \ \ \ \ \ \ \ \ \ \ \ \ \ \ \ \ \ \ \ \
\ \ \ \ \ \ \ \ \ \ \ \ \ \ \ \ \ \ \ \ \ \ \ \ \ \ \ \ \ \ \ \ \ \ \ \ \ \
\ \ \ \ \ \ \ \ \ \ \ \ \ \ \ \ \ \ \ \ \ \ \ \ \ \ \ \ \ \ \ \ \ \ \ \ \ \
\ \ \ \ \ \ \ \ \ \ \ \ \ \ \ \ \ \ \ \ \ \ \ \ \ \ \ \ \ \ \ \ \ \ \ \ \ \
\ \ \ \ \ \ \ \ \ \ \ \ \ \ \ \ \ \ \ \ \ \ \ \ \ \ \ \ \ \ \ \ \ \ \ \ \ \
\ \ \ \ \ \ \ \ \ \ \ \ \ \ \ \ \ \ \ \ \ \ \ \ \ \ \ \ \ \ \ \ \ \ \ \ \ \
\ \ \ \ \ \ \ \ \ \ \ \ \ \ \ \ \ \ \ \ \ \ \ \ \ \ \ \ \ \ \ \ \ \ \ \ \ \
\ \ \ \ \ \ \ \ \ \ \ \ \ \ \ \ \ \ \ \ \ \ \ \ \ \ \ \ \ \ \ \ \ \ \ \ \ \
\ \ \ \ \ \ \ \ \ \ \ \ \ \ \ \ \ \ \ \ \ \ \ \ \ \ \ \ \ \ \ \ \ \ \ \ \ \
\ \ \ \ \ \ \ \ \ \ \ \ \ \ \ \ \ \ \ \ \ \ \ \ \ \ \ \ \ \ \ \ \ \ \ \ \ \
\ \ \ \ \ \ \ \ \ \ \ \ \ \ \ \ \ \ \ \ \ \ \ \ \ \ \ \ \ \ \ \ \ \ \ \ \ \
\ \ \ \ \ \ \ \ \ \ \ \ \ \ \ \ \ \ \ \ \ \ \ \ \ \ \ \ \ \ \ \ \ \ \ \ \ \
\ \ \ \ \ \ \ \ \ \ \ \ \ \ \ \ \ \ \ \ \ \ \ \ \ \ \ \ \ \ \ \ \ \ \ \ \ \
\ \ \ \ \ \ \ \ \ \ \ \ \ \ \ \ \ \ \ \ \ \ \ \ \ \ \ \ \ \ \ \ \ \ \ \ \ \
\ \ \ \ \ \ \ \ \ \ \ \ \ \ \ \ \ \ \ \ \ \ \ \ \ \ \ \ \ \ \ \ \ \ \ \ \ \
\ \ \ \ \ \ \ \ \ \ \ \ \ \ \ \ \ \ \ \ \ \ \ \ \ \ \ \ \ \ \ \ \ \ \ \ \ \
\ \ \ \ \ \ \ \ \ \ \ \ \ \ \ \ \ \ \ \ \ \ \ \ \ \ \ \ \ \ \ \ \ \ \ \ \ \
\ \ \ \ \ \ \ \ \ \ \ \ \ \ \ \ \ \ \ \ \ \ \ \ \ \ \ \ \ \ \ \ \ \ \ \ \ \
\ \ \ \ \ \ \ \ \ \ \ \ \ \ \ \ \ \ \ \ \ \ \ \ \ \ \ \ \ \ \ \ \ \ \ \ \ \
\ \ \ \ \ \ \ \ \ \ \ \ \ \ \ \ \ \ \ \ \ \ \ \ \ \ \ \ \ \ \ \ \ \ \ \ \ \
\ \ \ \ \ \ \ \ \ \ \ \ \ \ \ \ \ \ \ \ \ \ \ \ \ \ \ \ \ \ \ \ \ \ \ \ \ \
\ \ \ \ \ \ \ \ \ \ \ \ \ \ \ \ \ \ \ \ \ \ \ \ \ \ \ \ \ \ \ \ \ \ \ \ \ \
\ \ \ \ \ \ \ \ \ \ \ \ \ \ \ \ \ \ \ \ \ \ \ \ \ \ \ \ \ \ \ \ \ \ \ \ \ \
\ \ \ \ \ \ \ \ \ \ \ \ \ \ \ \ \ \ \ \ \ \ \ \ \ \ \ \ \ \ \ \ \ \ \ \ \ \
\ \ \ \ \ \ \ \ \ \ \ \ \ \ \ \ \ \ \ \ \ \ \ \ \ \ \ \ \ \ \ \ \ \ \ \ \ \
\ \ \ \ \ \ \ \ \ \ \ \ \ \ \ \ \ \ \ \ \ \ \ \ \ \ \ \ \ \ \ \ \ \ \ \ \ \
\ \ \ \ \ \ \ \ \ \ \ \ \ \ \ \ \ \ \ \ \ \ \ \ \ \ \ \ \ \ \ \ \ \ \ \ \ \
\ \ \ \ \ \ \ \ \ \ \ \ \ \ \ \ \ \ \ \ \ \ \ \ \ \ \ \ \ \ \ \ \ \ \ \ \ \
\ \ \ \ \ \ \ \ \ \ \ \ \ \ \ \ \ \ \ \ \ \ \ \ \ \ \ \ \ \ \ \ \ \ \ \ \ \
\ \ \ \ \ \ \ \ \ \ \ \ \ \ \ \ \ \ \ \ \ \ \ \ \ \ \ \ \ \ \ \ \ \ \ \ \ \
\ \ \ \ \ \ \ \ \ \ \ \ \ \ \ \ \ \ \ \ \ \ \ \ \ \ \ \ \ \ \ \ \ \ \ \ \ \
\ \ \ \ \ \ \ \ \ \ \ \ \ \ \ \ \ \ \ \ \ \ \ \ \ \ \ \ \ \ \ \ \ \ \ \ \ \
\ \ \ \ \ \ \ \ \ \ \ \ \ \ \ \ \ \ \ \ \ \ \ \ \ \ \ \ \ \ \ \ \ \ \ \ \ \
\ \ \ \ \ \ \ \ \ \ \ \ \ \ \ \ \ \ \ \ \ \ \ \ \ \ \ \ \ \ \ \ \ \ \ \ \ \
\ \ \ \ \ \ \ \ \ \ \ \ \ \ \ \ \ \ \ \ \ \ \ \ \ \ \ \ \ \ \ \ \ \ \ \ \ \
\ \ \ \ \ \ \ \ \ \ \ \ \ \ \ \ \ \ \ \ \ \ \ \ \ \ \ \ \ \ \ \ \ \ \ \ \ \
\ \ \ \ \ \ \ \ \ \ \ \ \ \ \ \ \ \ \ \ \ \ \ \ \ \ \ \ \ \ \ \ \ \ \ \ \ \
\ \ \ \ \ \ \ \ \ \ \ \ \ \ \ \ \ \ \ \ \ \ \ \ \ \ \ \ \ \ \ \ \ \ \ \ \ \
\ \ \ \ \ \ \ \ \ \ \ \ \ \ \ \ \ \ \ \ \ \ \ \ \ \ \ \ \ \ \ \ \ \ \ \ \ \
\ \ \ \ \ \ \ \ \ \ \ \ \ \ \ \ \ \ \ \ \ \ \ \ \ \ \ \ \ \ \ \ \ \ \ \ \ \
\ \ \ \ \ \ \ \ \ \ \ \ \ \ \ \ \ \ \ \ \ \ \ \ \ \ \ \ \ \ \ \ \ \ \ \ \ \
\ \ \ \ \ \ \ \ \ \ \ \ \ \ \ \ \ \ \ \ \ \ \ \ \ \ \ \ \ \ \ \ \ \ \ \ \ \
\ \ \ \ \ \ \ \ \ \ \ \ \ \ \ \ \ \ \ \ \ \ \ \ \ \ \ \ \ \ \ \ \ \ \ \ \ \
\ \ \ \ \ \ \ \ \ \ \ \ \ \ \ \ \ \ \ \ \ \ \ \ \ \ \ \ \ \ \ \ \ \ \ \ \ \
\ \ \ \ \ \ \ \ \ \ \ \ \ \ \ \ \ \ \ \ \ \ \ \ \ \ \ \ \ \ \ \ \ \ \ \ \ \
\ \ \ \ \ \ \ \ \ \ \ \ \ \ \ \ \ \ \ \ \ \ \ \ \ \ \ \ \ \ \ \ \ \ \ \ \ \
\ \ \ \ \ \ \ \ \ \ \ \ \ \ \ \ \ \ \ \ \ \ \ \ \ \ \ \ \ \ \ \ \ \ \ \ \ \
\ \ \ \ \ \ \ \ \ \ \ \ \ \ \ \ \ \ \ \ \ \ \ \ \ \ \ \ \ \ \ \ \ \ \ \ \ \
\ \ \ \ \ \ \ \ \ \ \ \ \ \ \ \ \ \ \ \ \ \ \ \ \ \ \ \ \ \ \ \ \ \ \ \ \ \
\ \ \ \ \ \ \ \ \ \ \ \ \ \ \ \ \ \ \ \ \ \ \ \ \ \ \ \ \ \ \ \ \ \ \ \ \ \
\ \ \ \ \ \ \ \ \ \ \ \ \ \ \ \ \ \ \ \ \ \ \ \ \ \ \ \ \ \ \ \ \ \ \ \ \ \
\ \ \ \ \ \ \ \ \ \ \ \ \ \ \ \ \ \ \ \ \ \ \ \ \ \ \ \ \ \ \ \ \ \ \ \ \ \
\ \ \ \ \ \ \ \ \ \ \ \ \ \ \ \ \ \ \ \ \ \ \ \ \ \ \ \ \ \ \ \ \ \ \ \ \ \
\ \ \ \ \ \ \ \ \ \ \ \ \ \ \ \ \ \ \ \ \ \ \ \ \ \ \ \ \ \ \ \ \ \ \ \ \ \
\ \ \ \ \ \ \ \ \ \ \ \ \ \ \ \ \ \ \ \ \ \ \ \ \ \ \ \ \ \ \ \ \ \ \ \ \ \
\ \ \ \ \ \ \ \ \ \ \ \ \ \ \ \ \ \ \ \ \ \ \ \ \ \ \ \ \ \ \ \ \ \ \ \ \ \
\ \ \ \ \ \ \ \ \ \ \ \ \ \ \ \ \ \ \ \ \ \ \ \ \ \ \ \ \ \ \ \ \ \ \ \ \ \
\ \ \ \ \ \ \ \ \ \ \ \ \ \ \ \ \ \ \ \ \ \ \ \ \ \ \ \ \ \ \ \ \ \ \ \ \ \
\ \ \ \ \ \ \ \ \ \ \ \ \ \ \ \ \ \ \ \ \ \ \ \ \ \ \ \ \ \ \ \ \ \ \ \ \ \
\ \ \ \ \ \ \ \ \ \ \ \ \ \ \ \ \ \ \ \ \ \ \ \ \ \ \ \ \ \ \ \ \ \ \ \ \ \
\ \ \ \ \ \ \ \ \ \ \ \ \ \ \ \ \ \ \ \ \ \ \ \ \ \ \ \ \ \ \ \ \ \ \ \ \ \
\ \ \ \ \ \ \ \ \ \ \ \ \ \ \ \ \ \ \ \ \ \ \ \ \ \ \ \ \ \ \ \ \ \ \ \ \ \
\ \ \ \ \ \ \ \ \ \ \ \ \ \ \ \ \ \ \ \ \ \ \ \ \ \ \ \ \ \ \ \ \ \ \ \ \ \
\ \ \ \ \ \ \ \ \ \ \ \ \ \ \ \ \ \ \ \ \ \ \ \ \ \ \ \ \ \ \ \ \ \ \ \ \ \
\ \ \ \ \ \ \ \ \ \ \ \ \ \ \ \ \ \ \ \ \ \ \ \ \ \ \ \ \ \ \ \ \ \ \ \ \ \
\ \ \ \ \ \ \ \ \ \ \ \ \ \ \ \ \ \ \ \ \ \ \ \ \ \ \ \ \ \ \ \ \ \ \ \ \ \
\ \ \ \ \ \ \ \ \ \ \ \ \ \ \ \ \ \ \ \ \ \ \ \ \ \ \ \ \ \ \ \ \ \ \ \ \ \
\ \ \ \ \ \ \ \ \ \ \ \ \ \ \ \ \ \ \ \ \ \ \ \ \ \ \ \ \ \ \ \ \ \ \ \ \ \
\ \ \ \ \ \ \ \ \ \ \ \ \ \ \ \ \ \ \ \ \ \ \ \ \ \ \ \ \ \ \ \ \ \ \ \ \ \
\ \ \ \ \ \ \ \ \ \ \ \ \ \ \ 

This paper predicts some quarks, baryons and mesons shown in the following
list:

\begin{tabular}{|l|l|l|l|l|l|l|l|l|}
\hline
{\small q}$_{i}${\small (m)} & {\small q}$_{j}$ & {\small q}$_{k}$ & {\small %
Baryon} & {\small *} & {\small q}$_{i}${\small (m)}$\overline{\text{q}_{j}%
\text{(m)}}$ & {\small Meson} & {\small *} & {\small q}$_{i}${\small (m)}$%
\overline{\text{{\small q}}_{i}\text{{\small (m)}}}^{\$}$ \\ \hline
$\text{u}_{C}${\small (6073)} & {\small u} & {\small d} & $\Lambda _{C}$%
{\small (6699)} & {\small *} & $\text{u}_{C}^{1}${\small (6073)}$\overline{%
\text{{\small q}}_{N}^{0}\text{{\small (313)}}}$ & {\small D(6231)} & 
{\small *} & $\psi ${\small (10509)}$^{{\small \#}}$ \\ \hline
$\text{d}_{b}^{1}${\small (9333)} & {\small u} & {\small d} & {\small B(9959)%
} & {\small *} & {\small q}$_{N}^{0}${\small (313)}$\overline{\text{d}%
_{b}^{1}\text{{\small (9333)}}}$ & {\small B(9502)} & {\small *} & $\Upsilon 
${\small (17868)} \\ \hline
$\text{d}_{S}\text{({\small 773})}$ & {\small u} & {\small d} & $\Lambda $%
{\small (1399)}$^{\#}$ & {\small *} & {\small q}$_{N}^{0}${\small (313)}$%
\overline{\text{d}_{S}^{-1}\text{{\small (773)}}}$ & {\small K(916)}$^{\#}$
& {\small *} & $\phi ${\small (1041)}$^{{\small \#}}$ \\ \hline
$\text{d}_{S}\text{({\small 1933})}$ & {\small u} & {\small d} & $\Lambda $%
{\small (2559)}$^{\#}$ & {\small *} & {\small q}$_{N}^{0}${\small (313)}$%
\overline{\text{d}_{S}^{1}\text{{\small (1933)}}}$ & {\small K(2076)} & 
{\small *} & $\eta ${\small (3005)}$^{{\small \#}}$ \\ \hline
$\text{d}_{S}\text{({\small 3753})}$ & {\small u} & {\small d} & $\Lambda $%
{\small (4379)} & {\small *} & {\small q}$_{N}^{0}${\small (313)}$\overline{%
\text{d}_{S}^{-1}\text{{\small (3753)}}}$ & {\small K(3876)} & {\small *} & $%
\eta ${\small (5472)} \\ \hline
$\text{d}_{S}\text{({\small 9613})}$ & {\small u} & {\small d} & $\Lambda $%
{\small (10239)} & {\small *} & {\small q}$_{N}^{0}${\small (313)}$\overline{%
\text{{\small d}}_{S}^{-1}\text{{\small (9613)}}}$ & {\small K(9781)} & 
{\small *} & $\eta ${\small (18133)} \\ \hline
\end{tabular}

\qquad\ $^{\$}${\small The last column shows the mesons of the pair quarks [q%
}$_{i}${\small (m)}$\overline{\text{{\small q}}_{i}\text{{\small (m)}}}$%
{\small ] that the}

{\small \ \ \ \ \ \ \ \ \ \ \ \ q}$_{i}${\small (m) is in the first column.}

$\ \ \ \ \ \ \ \ \ \ \Lambda ^{0}${\small (1399)}$^{\#}${\small \ \ \ \ \
[experimental }$\Lambda ^{0}${\small (1406) with (}$\frac{\Delta M}{M}$%
{\small \%) = 0.5\%],}

{\small \ \ \ \ \ \ \ \ \ \ \ \ }$\Lambda ${\small (2559)}$^{\#}${\small \ \
\ \ \ \ \ [experimental }$\Lambda ${\small (2585)}$^{\ast \ast }${\small \
with (}$\frac{\Delta M}{M}${\small \%) = 1.0\%],}

{\small \ \ \ \ \ \ \ \ \ \ \ \ K(916)}$^{\#}${\small \ \ \ \ \ \ \ \ \ \
[experimental K}$^{\ast }${\small (892) with (}$\frac{\Delta M}{M}${\small %
\%) = 2.7\%],}

{\small \ \ \ \ \ }$\ \ \ \ \ \ \eta ${\small (1041)}$^{\#}${\small \ \ \ \
\ \ \ \ [experimental }$\phi ${\small (1020) with (}$\frac{\Delta M}{M}$%
{\small \%) = 2\%],}

{\small \ \ \ \ \ }$\ \ \ \ \ \ \eta ${\small (3005)}$^{\#}${\small \ \ \ \
\ \ \ \ [experimental }$\eta _{c}${\small (2980) with (}$\frac{\Delta M}{M}$%
{\small \%) = 0.8\%].}

It is very important to pay attention to the $\Upsilon $(3S)-meson (mass m =
10,355.2 $\pm $ 0.4 Mev, full width $\Gamma $\ = 26.3 $\pm $ 3.5 kev). We
compare the mesons J/$\psi $(3097), $\Upsilon $(9460) and $\Upsilon $(10355)
shown as follow list

\begin{tabular}{l}
u$_{C}^{1}$(1753)$\overline{\text{u}_{C}^{1}\text{(1753)}}$ = J/$\psi $%
(3069) \ [J/$\psi $(3096.916$\pm $0.011), $\Gamma $ = 91.0 $\pm $ 3.2kev] \\ 
d$_{b}^{1}$(4913)$\overline{\text{d}_{b}^{1}\text{(4913)}}$ = $\Upsilon $%
(9389) \ \ \ \ \ [$\Upsilon $(9460.30$\pm $0.26), $\ \ \ \ \ \ \ \Gamma $ =
53.0 $\pm $ 1.5kev] \\ 
u$_{C}^{1}$(6073)$\overline{\text{u}_{C}^{1}\text{(6073)}}$ = $\psi $(10509)
\ \ [$\Upsilon $(10.355.2 $\pm $ 0.4), $\ \ \ \ \ \Gamma $\ = 26.3 $\pm $
3.5 kev]%
\end{tabular}

$\Upsilon $(3S) has more than three times larger of a mass than J/$\psi $%
(1S)\ (m = 3096.916 $\pm $ 0.011 Mev)\ and more than three times longer of a
lifetime than J/$\psi $(1S)\ (full width $\Gamma $ = 91.0 $\pm $ \ 3.2 kev).
It is well known that the discovery\ of J/$\psi $(1S)\ is also the discovery
of charmed\ quark c (u$_{c}$(1753)) and that the discovery\ of $\Upsilon $%
(9460)\ is also the discovery of bottom\ quark b (d$_{b}$(4913)). Similarly
the discovery of $\Upsilon $(3S) will be the discovery of a very important
new quark---the u$_{C}$(6073)-quark.

\ \ \ \ \ \ \ \ \ \ \ \ \ \ \ \ \ \ \ \ \ \ \ \ \ \ \ \ \ \ \ \ \ \ \ \ \ 

\textbf{9 Discussion\ }

\ \ \ \ \ \ \ \ \ \ \ \ \ \ \ \ \ \ \ \ \ \ \ \ \ 

1). From the low energy free wave motion of a excited elementary quark $%
\epsilon $ with a continuous energy spectrum \{$\mathbb{E}$ = V +$\frac{%
\hslash ^{2}}{2m}$[(k$_{1}$)$^{2}$+(k$_{2}$)$^{2}$+(k$_{3}$)$^{2}$]\}, using
the three step quantization, we obtain a new energy formula \{$\mathbb{E}$($%
\vec{k}$,$\vec{n}$) =313 + $\Delta $ + $\alpha $[(n$_{1}$-$\xi $)$^{2}$+(n$%
_{2}$-$\eta $)$^{2}$+(n$_{3}$-$\zeta $)$^{2}$]\} with quantized $\vec{n}$
values (\ref{nnn}) and $\vec{k}$ values (\ref{Sym-Axes}). The energy (\ref%
{E(nk)}) with a $\overrightarrow{n}$\ = (n$_{1}$, n$_{2}$, n$_{3}$) of (\ref%
{nnn}) and a $\vec{k}$ = ($\xi $, $\eta $, $\varsigma $) of (\ref{Sym-Axes})
forms an energy band. If the free eigen wave function and eigen energy of
the Schr\"{o}dinger equation are first quantization, the three step
quantization is the \textquotedblleft second quantization\textquotedblright
. This \textquotedblleft second quantization\textquotedblright\ products new
quantum-- shout-lived quarks.

2). The rest masses and the intrinsic quantum numbers (I, S, C, B and Q) are
necessary for the standard model, but they cannot be deduced by the standard
model. Using (\ref{IsoSpin}) - (\ref{Rest Mass}), we deduce the rest masses
and intrinsic quantum numbers of quarks from the energy bands. The deduced
rest masses and quantum numbers of baryons and mesons from these masses and
numbers of quarks, are\textbf{\ }consistent with experimental results. This
is a strong support for the three step quantization.\ \ \ \ \ \ \ \ \ \ \ \
\ \ \ \ \ \ \ \ \ \ \ \ \ \ \ \ \ \ \ \ \ \ \ \ \ \ \ \ \ \ \ \ \ \ \ \ \ \
\ \ \ \ \ \ \ \ \ \ \ \ \ \ \ \ \ \ \ \ \ \ \ \ \ \ \ \ \ \ \ \ \ \ \ \ \ \
\ \ \ \ \ \ \ \ \ \ \ \ \ \ \ \ \ \ \ \ \ \ \ \ \ \ \ \ \ \ \ \ \ \ \ \ \ \
\ \ \ \ \ \ \ \ \ \ \ \ \ \ \ \ \ \ \ \ \ \ \ \ \ \ \ \ \ \ \ \ \ \ \ \ \ \
\ \ \ \ \ \ \ \ \ \ \ \ \ \ \ \ \ \ \ \ \ \ \ \ \ \ \ \ \ \ \ \ \ \ \ \ \ \
\ \ \ \ \ \ \ \ \ \ \ \ \ \ \ \ \ \ \ \ \ \ \ \ \ \ \ \ \ \ \ \ \ \ \ \ \ \
\ \ \ \ \ \ \ \ \ \ \ \ \ \ \ \ \ \ \ \ \ \ \ \ \ \ \ \ \ \ \ \ \ \ \ \ \ \
\ \ \ \ \ \ \ \ \ \ \ \ \ \ \ \ \ \ \ \ \ \ \ \ \ \ \ \ \ \ \ \ \ \ \ \ \ \
\ \ \ \ \ \ \ \ \ \ \ \ \ \ \ \ \ \ \ \ \ \ \ \ \ \ \ \ \ \ \ \ \ \ \ \ \ \
\ \ \ \ \ \ \ \ \ \ \ \ \ \ \ \ \ \ \ \ \ \ \ \ \ \ \ \ \ \ \ \ \ \ \ \ \ \
\ \ \ \ \ \ \ \ \ \ \ \ \ \ \ \ \ \ \ \ \ \ \ \ \ \ \ \ \ \ \ \ \ \ \ \ \ \
\ \ \ \ \ \ \ \ \ \ \ \ \ \ \ \ \ \ \ \ \ \ \ \ \ \ \ \ \ \ \ \ \ \ \ \ \ \
\ \ \ \ \ \ \ \ \ \ \ \ \ \ \ \ \ \ \ \ \ \ \ \ \ \ \ \ \ \ \ \ \ \ \ \ \ \
\ \ \ \ \ \ \ \ \ \ \ \ \ \ \ \ \ \ \ \ \ \ \ \ \ \ \ \ \ \ \ \ \ \ \ \ \ \
\ \ \ \ \ \ \ \ \ \ \ \ \ \ \ \ \ \ \ \ \ \ \ \ \ \ \ \ \ \ \ \ \ \ \ \ \ \
\ \ \ \ \ \ \ \ \ \ \ \ \ \ \ \ \ \ \ \ \ \ \ \ \ \ \ \ \ \ \ \ \ \ \ \ \ \
\ \ \ \ \ \ \ \ \ \ \ \ \ \ \ \ \ \ \ \ \ \ \ \ \ \ \ \ \ \ \ \ \ \ \ \ \ \
\ \ \ \ \ \ \ \ \ \ \ \ \ \ \ \ \ \ \ \ \ \ \ \ \ \ \ \ \ \ \ \ \ \ \ \ \ \
\ \ \ \ \ \ \ \ \ \ \ \ \ \ \ \ \ \ \ \ \ \ \ \ \ \ \ \ \ \ \ \ \ \ \ \ \ \
\ \ \ \ \ \ \ \ \ \ \ \ \ \ \ \ \ \ \ \ \ \ \ \ \ \ \ \ \ \ \ \ \ \ \ \ \ \
\ \ \ \ \ \ \ \ \ \ \ \ \ \ \ \ \ \ \ \ \ \ \ \ \ \ \ \ \ \ \ \ \ \ \ \ \ \
\ \ \ \ \ \ \ \ \ \ \ \ \ \ \ \ \ \ \ \ \ \ \ \ \ \ \ \ \ \ \ \ \ \ \ \ \ \
\ \ \ \ \ \ \ \ \ \ \ \ \ \ \ \ \ \ \ \ \ \ \ \ \ \ \ \ \ \ \ \ \ \ \ \ \ \
\ \ \ \ \ \ \ \ \ \ \ \ \ \ \ \ \ \ \ \ \ \ \ \ \ \ \ \ \ \ \ \ \ \ \ \ \ \
\ \ \ \ \ \ \ \ \ \ \ \ \ \ \ \ \ \ \ \ \ \ \ \ \ \ \ \ \ \ \ \ \ \ \ \ \ \
\ \ \ \ \ \ \ \ \ \ \ \ \ \ \ \ \ \ \ \ \ \ \ \ \ \ \ \ \ \ \ \ \ \ \ \ \ \
\ \ \ \ \ \ \ \ \ \ \ \ \ \ \ \ \ \ \ \ \ \ \ \ \ \ \ \ \ \ \ \ \ \ \ \ \ \
\ \ \ \ \ \ \ \ \ \ \ \ \ \ \ \ \ \ \ \ \ \ \ \ \ \ \ \ \ \ \ \ \ \ \ \ \ \
\ \ \ \ \ \ \ \ \ \ \ \ \ \ \ \ \ \ \ \ \ \ \ \ \ \ \ \ \ \ \ \ \ \ \ \ \ \
\ \ \ \ \ \ \ \ \ \ \ \ \ \ \ \ \ \ \ \ \ \ \ \ \ \ \ \ \ \ \ \ \ \ \ \ \ \
\ \ \ \ \ \ \ \ \ \ \ \ \ \ \ \ \ \ \ \ \ \ \ \ \ \ \ \ \ \ \ \ \ \ \ \ \ \
\ \ \ \ \ \ \ \ \ \ \ \ \ \ \ \ \ \ \ \ \ \ 

3). The five quarks of the current Quark Model correspond to the five
deduced ground quarks [u$\leftrightarrow $u(313), d$\leftrightarrow $d(313),
s$\leftrightarrow $d$_{s}$(493), c$\leftrightarrow $u$_{c}$(1753) and b$%
\leftrightarrow $d$_{b}$(4913)] (see Table 11 of \cite{0502091})). The
current Quark Model uses only these five quarks to explain baryons and
mesons. In early times, this was reasonable, natural and useful. Today,
however, it is not reasonable that physicists use only these five current
quarks since physicists have discovered many high energy baryons and mesons
that are composed of more high energy quarks.

4). The energy band excited quarks u(313) with $\overrightarrow{n}$ = (0, 0,
0) in Table 3 and d(313) with $\overrightarrow{n}$ = (0, 0, 0) in Table 4
will be short-lived quarks. They are, however, lowest energy quarks. Since
there is no lower energy position that they can decay into, they are not
short-lived quarks. Because they have the same rest mass and intrinsic
quantum numbers as the free excited quarks u(313) and d(313), they cannot be
distinguished from the free excited quarks u(313) and d(313) by experiments.
The u(313) and d(313) with $\overrightarrow{n}$ = (0, 0, 0) will be covered
up by free excited u(313) and d(313) in experiments. Therefore, we can omit
u(313) and d(313) with $\overrightarrow{n}$ = (0, 0, 0) since the
probability that they are produced much small than the free excited
u(313)-quark and d(313)-quark. There are only long-lived and free excited
the u(313)-quark and the d(313)-quark in both theory and experiments. \ 

5). The fact that physicists have not found any free quark shows that the
binding energies are very large. The baryon binding energy -3$\Delta $
(meson - 2$\Delta $ ) is a phenomenological approximation of the color's
strong interaction energy in a baryon (a meson). The binding energy -3$%
\Delta $ (-2$\Delta $) is always cancelled by the corresponding parts 3$%
\Delta $ of the rest masses of the three quarks in a baryon (2$\Delta $ of
the quark and antiquark in a meson). Thus we can omit the binding energy -3$%
\Delta $ (-2$\Delta $) and the corresponding rest mass parts 3$\Delta $ (2$%
\Delta $) of the quarks when we deduce rest masses of baryons (mesons). This
effect makes it appear as if there is no strong binding energy in baryons
(mesons).

\ \ \ \ \ \ \ \ \ \ \ \ \ \ \ \ \ \ \ \ \ \ \ \ \ \ \ \ \ \ \ \ \ \ 

\textbf{10 Conclusions\ \ }\ \ \ \ \ \ \ \ \ \ \ \ \ \ \ \ \ \ \ \ \ \ \ \ \
\ \ \ \ \ \ \ 

\ \ \ \ \ \ \ \ \ \ \ \ \ \ \ \ \ \ \ \ \ \ \ \ \ \ \ \ \ \ \ \ \ \ \ 

1). There is only one elementary quark family $\epsilon $ with three colors
and two isospin states ($\epsilon _{u}$ with I$_{Z}$ = $\frac{1}{2}$ and Q =
+$\frac{2}{3}$, $\epsilon _{d}$ with I$_{Z}$ = $\frac{-1}{2}$ and Q = -$%
\frac{1}{3}$) for each color. Thus there are six Fermi (s = $\frac{1}{2}$)
elementary quarks with S = C = B = 0 in the vacuum. $\epsilon _{u}$ and $%
\epsilon _{d}$ have SU(2) symmetries.

2). All quarks inside hadrons are the excited states of the elementary quark 
$\epsilon $. There are two types of excited states: free excited states and
energy band excited states. The free excited states are only the u-quark and
the d-quark. They are long-lived quarks. The energy band excited states are
the short-lived quarks, such as d$_{s}$(493), d$_{s}$(773), u$_{c}$(1753)
and d$_{b}$(4913).

3). Since all quarks inside hadrons are excited states of the same
elementary quark $\epsilon $, all quarks (m \TEXTsymbol{>} 313 Mev) will
eventually decay into the q$_{N}$(313)-quark [(u(313) and d(313)].

4). The three step quantization is a someway \textquotedblleft second
quantization\textquotedblright\ that products the short-lived quarks.

5). There is a large binding energy -3$\Delta $ (or -2$\Delta $) among three
quarks in a baryon (or between the quark and the antiquark in a meson). It
may be a possible foundation for the quark confinement.

6). We have deduced the rest masses and intrinsic quantum numbers of quarks
(Table 3 and 4), baryons (Table 6) and mesons (Table 7) using the three step
quantization method and phenomenological formulae. The deduced intrinsic
quantum numbers of baryons and mesons match the experimental results \cite%
{Baryon04} and \cite{Meson04} exactly, while the deduced rest masses of the
baryons and the mesons are consistent with more than 98\% of experimental
results \cite{Baryon04} and \cite{Meson04}.

7). The current Quark Model is the five ground quark approximation of a more
fundamental model.

8). This paper predict some new quarks [$\text{u}_{C}$(6073) , $\text{d}_{b}$%
(9333) and d$_{S}$(773)], baryons [$\Lambda _{C}$(6699){\small \ and }%
B(9959)] and mesons [D(6231), B(9502) and K(3876)].

\ \ \ \ \ \ \ \ \ \ \ \ \ \ \ \ \ \ \ \ \ \ \ \ \ \ \ \ \ \ \ \ \ \ \ \ \ \
\ \ \ \ 

\begin{center}
\bigskip \textbf{Acknowledgments}
\end{center}

I sincerely thank Professor Robert L. Anderson for his valuable advice. I
acknowledge\textbf{\ }my indebtedness to Professor D. P. Landau for his help
also. I would like to express my heartfelt gratitude to Dr. Xin Yu for
checking the calculations. I sincerely thank Professor Yong-Shi Wu for his
important advice and help. I thank Professor Wei-Kun Ge for his support and
help. I sincerely thank Professor Kang-Jie Shi for his advice.

\bigskip\ \ \ \ \ \ \ \ 

\end{document}